\documentstyle[epsf]{mn}

\newif\ifAMStwofonts

\ifoldfss
  \ifCUPmtlplainloaded \else
    \NewTextAlphabet{textbfit} {cmbxti10} {}
    \NewTextAlphabet{textbfss} {cmssbx10} {}
    \NewMathAlphabet{mathbfit} {cmbxti10} {} 
    \NewMathAlphabet{mathbfss} {cmssbx10} {} 
  \fi
  \ifAMStwofonts
    \ifCUPmtlplainloaded \else
      \NewSymbolFont{upmath} {eurm10}
      \NewSymbolFont{AMSa} {msam10}
      \NewMathSymbol{\upi}     {0}{upmath}{19}
      \NewMathSymbol{\umu}     {0}{upmath}{16}
      \NewMathSymbol{\upartial}{0}{upmath}{40}
      \NewMathSymbol{\leqslant}{3}{AMSa}{36}
      \NewMathSymbol{\geqslant}{3}{AMSa}{3E}

    \fi
  \fi
\fi 

\ifnfssone
  \newmathalphabet{\mathit}
  \addtoversion{normal}{\mathit}{cmr}{m}{it}
  \addtoversion{bold}{\mathit}{cmr}{bx}{it}
  \newmathalphabet{\mathbfit} 
  \addtoversion{normal}{\mathbfit}{cmr}{bx}{it}
  \addtoversion{bold}{\mathbfit}{cmr}{bx}{it}
  \newmathalphabet{\mathbfss} 
  \addtoversion{normal}{\mathbfss}{cmss}{bx}{n}
  \addtoversion{bold}{\mathbfss}{cmss}{bx}{n}
  \ifAMStwofonts
    \ifCUPmtlplainloaded \else
      \UseAMStwoboldmath
      \makeatletter
      \new@mathgroup\upmath@group
      \define@mathgroup\mv@normal\upmath@group{eur}{m}{n}
      \define@mathgroup\mv@bold\upmath@group{eur}{b}{n}
      \edef\UPM{\hexnumber\upmath@group}
      \new@mathgroup\amsa@group
      \define@mathgroup\mv@normal\amsa@group{msa}{m}{n}
      \define@mathgroup\mv@bold\amsa@group{msa}{m}{n}
      \edef\AMSa{\hexnumber\amsa@group}
      \makeatother
      \mathchardef\upi="0\UPM19
      \mathchardef\umu="0\UPM16
      \mathchardef\upartial="0\UPM40
      \mathchardef\leqslant="3\AMSa36
      \mathchardef\geqslant="3\AMSa3E
    \fi
  \fi
\fi 

\ifnfsstwo
  \DeclareMathAlphabet{\mathbfit}{OT1}{cmr}{bx}{it}
  \SetMathAlphabet\mathbfit{bold}{OT1}{cmr}{bx}{it}
  \DeclareMathAlphabet{\mathbfss}{OT1}{cmss}{bx}{n}
  \SetMathAlphabet\mathbfss{bold}{OT1}{cmss}{bx}{n}
  \ifAMStwofonts
    \ifCUPmtlplainloaded \else
      \DeclareSymbolFont{UPM}{U}{eur}{m}{n}
      \SetSymbolFont{UPM}{bold}{U}{eur}{b}{n}
      \DeclareSymbolFont{AMSa}{U}{msa}{m}{n}
      \DeclareMathSymbol{\upi}{0}{UPM}{"19}
      \DeclareMathSymbol{\umu}{0}{UPM}{"16}
      \DeclareMathSymbol{\upartial}{0}{UPM}{"40}
      \DeclareMathSymbol{\leqslant}{3}{AMSa}{"36}
      \DeclareMathSymbol{\geqslant}{3}{AMSa}{"3E}
    \fi
  \fi
\fi 

\ifCUPmtlplainloaded \else
  \ifAMStwofonts \else 
    \def\upi{\pi}
    \def\umu{\mu}
    \def\upartial{\partial}
  \fi
\fi

\title{Multiwavelength observations of the M\,15 intermediate velocity cloud}

\author[J.V. Smoker et al.]
         {J.~V. Smoker$^{1}$, L.~M. Haffner$^{2}$, F.~P. Keenan$^{1}$, 
          R.~D. Davies$^{3}$, D. Pollacco$^{1}$  \\     
       $^{1}$Astrophysics and Planetary Science Division,
            Department of Pure and Applied Physics,
            The Queen's University of Belfast, \\
            University Road, Belfast, BT7 1NN, 
            U.K. \\
     $^{2}$Dept of Astronomy,
           University of Wisconsin,
           5534 Sterling Hall,
           475 North Charter Street,
           Madison, WI 53706,
           U.S.A.        \\
    $^{3}$Jodrell Bank Observatory,
          University of Manchester,
          near Macclesfield,
          Cheshire,
          SK11 9DL,
          U.K.            \\
}

\date{Accepted 
      Received 
      in original form }

\pubyear{}

\begin{document}
\maketitle

\label{firstpage}

\begin{abstract}

We present Westerbork Synthesis Radio Telescope H\,{\sc i} images, Lovell 
telescope multibeam H\,{\sc i} wide-field mapping, William Herschel Telescope longslit 
echelle Ca\,{\sc ii} observations, Wisconsin H$\alpha$ Mapper (WHAM) facility 
images, and IRAS ISSA 60 and 100 micron coadded images towards the intermediate velocity 
cloud (IVC) at +70 km\,s$^{-1}$, located in the general direction of the M\,15 globular cluster. 
When combined with previously-published Arecibo data, the H\,{\sc i} 
gas in the IVC is found to be clumpy, with a peak 
H\,{\sc i} column density of $\sim$ 1.5 $\times$ 10$^{20}$ cm$^{-2}$, inferred volume 
density (assuming spherical symmetry) of $\sim$ 
24 cm$^{-3}$ / $D$ (kpc), and maximum brightness temperature at a resolution of 
$81^{\prime\prime} \times 14^{\prime\prime}$ of $\sim$ 14 K. The major axis of 
this part of the IVC lies approximately parallel with the Galactic plane, as does the low 
velocity H\,{\sc i} gas and IRAS emission. The H\,{\sc i} gas in the cloud is  
warm, with a minimum value of the full width half maximum (FWHM) velocity 
width of 5 km\,s$^{-1}$ corresponding to a kinetic temperature, in the absence of
turbulence, of $\sim$ 540 K. From the H\,{\sc i} data, there are indications 
of two-component velocity structure. Similarly, the Ca\,{\sc ii} spectra, of 
resolution 7 km\,s$^{-1}$, also show {\em tentative} evidence of velocity structure, 
perhaps indicative of cloudlets. Assuming there are no unresolved narrow-velocity 
components, the mean values of log$_{10}$(N(Ca\,{\sc ii} K) cm$^{-2}$) $\sim$ 12.0 and 
Ca\,{\sc ii}/H\,{\sc i} $\sim$ 2.5$\times$10$^{-8}$ are typical of observations of high 
Galactic latitude clouds. This compares with a value of Ca\,{\sc ii}/H\,{\sc i} $>$ 10$^{-6}$ for IVC absorption 
towards HD 203664, a halo star of distance 3 kpc, some 3.1 degrees from the 
main M\,15 IVC condensation. The main IVC condensation is detected by WHAM 
in H$\alpha$ with central LSR velocities of $\sim$ 60$-$70 km\,s$^{-1}$, and intensities 
uncorrected for Galactic extinction of up to 1.3 Rayleigh, indicating that the gas is 
partially ionised. The FWHM values of the H$\alpha$ IVC component, at a resolution of 
1 degree, exceed 30 km\,s$^{-1}$. This is some 10 km\,s$^{-1}$ larger than the corresponding 
H\,{\sc i} value at similar resolution, and indicates that the two components may not be mixed. 
However, the spatial and velocity coincidence of the H$\alpha$ and H\,{\sc i} peaks in 
emission towards the main IVC component is qualitatively good. If the H$\alpha$ 
emission is caused solely by photoionisation, the Lyman continuum flux towards 
the main IVC condensation is $\sim$ 2.7$\times$10$^{6}$ photons cm$^{-2}$ s$^{-1}$. 
There is not a corresponding IVC H$\alpha$ detection towards the halo star HD 203664 
at velocities exceeding $\sim$ 60 km\,s$^{-1}$. Finally, both the 
60 and 100 micron IRAS images show spatial coincidence, over a 0.675$^{\circ} 
\times$0.625$^{\circ}$ field, with both low and intermediate velocity H\,{\sc i} gas 
(previously observed with the Arecibo telescope), indicating 
that the IVC may contain dust. Both the H$\alpha$ and tentative IRAS detections 
discriminate this IVC from High Velocity Clouds although the H\,{\sc i} properties 
do not. When combined with the H\,{\sc i} and optical results, these data point to a 
Galactic origin for at least parts of this IVC. 

\end{abstract}

\begin{keywords}
 ISM: general --
 ISM: clouds --
 ISM: individual objects: Complex gp --
 ISM: structure --
 globular clusters: individual: M\,15 --
 radio lines: ISM.
\end{keywords}

\section{Introduction}

The study of intermediate velocity clouds (IVCs) remains one of the most
challenging in contemporary Galactic astronomy, with several issues concerning 
IVCs remaining unresolved. These include, but are not limited to, 
the method of their formation, their relationship (if any) with high velocity 
clouds (HVCs), and the question as to whether IVCs are sites of star formation 
in the halo of the Galaxy (Kuntz \& Danly 1996; Christodoulou, Tohline 
\& Keenan 1997; Ivezic \& Christodoulou 1997). This latter possibility is underpinned 
by the fact that within the Galactic halo, there exists a population of early B-type stars 
whose velocities, ages and distances from the Galactic plane ($z$) 
are incompatible with them being formed within the disc. A possible site for their 
formation is IVCs/HVCs via cloud-cloud collisions and subsequent compression of the gas 
(Dyson \& Hartquist 1983). Such collisions are thought to be a viable 
star formation mechanism within at least the discs of galaxies, albeit where the 
gas density and cloud-cloud collision rates are somewhat higher than inferred 
in IVCs/HVCs (Tan 2000).

The solution to both the star formation question and also any possible relationship
between HVCs and IVCs requires both the analysis of aggregate parameters of well-defined 
samples of IVCs and HVCs, and also more detailed studies of individual objects. In this 
paper we report on radio H\,{\sc i} aperture synthesis, H\,{\sc i} multibeam wide field 
mapping, longslit Ca\,{\sc ii} observations, Wisconsin H$\alpha$ Mapper (WHAM) facility 
images, and IRAS sky-survey archive data retrieval 
towards a particular IVC located in the general 
direction of the M\,15 globular cluster (RA=21$^{\rm h}$ 29$^{\rm m}$ 58.29$^{\rm s}$, 
Dec=+12$^{\circ}$ 10$^{\prime}$ 00.5$^{\prime\prime}$ (J2000); 
$l$=65.01$^{\circ}$, $b$=--27.31$^{\circ}$). These observations are amongst the 
first H\,{\sc i} synthesis data to be taken of positive-velocity IVCs, which 
remain poorly-studied as a group of objects. 

The M\,15 H\,{\sc i} cloud lies at a velocity of $\sim$ +70 km\,s$^{-1}$ in the 
dynamical Local Standard of Rest (Allen 1973); its distance tentatively lies 
between $\sim$ 0.8--3 kpc (Little et al. 1994; Smoker et al. 2001a). The upper 
distance limit is gleaned from the fact that IVC absorption at $\sim$+70 to +80 km\,s$^{-1}$ 
is observed in the spectrum of HD 203664, a halo star of distance 3 kpc and $\sim$ 3.1 
degrees from M\,15 (Little et al. 1994), combined with the detection of IV H\,{\sc i} 
approximately mid-way between the M\,15 IVC and HD 203664. 

The deviation velocity of the M\,15 IVC at such a mid-Galactic 
latitude puts it in on the borderline between the normal definitions for 
intermediate and high velocity clouds (c.f. Fig. 1 of Wakker 1991), although in 
common with Sembach (1995) and Kennedy et al. (1998) here we classify it as an IVC. 
The line-of-sight position of the M\,15 IVC is between the negative-velocity Local 
Group barycentre cloud Complex G and the Galactic centre clouds (Fig. 8 of Blitz 
et al. 1999), hence the M\,15 IVC is a part of IVC Complex gp (Wakker 2001).

Previous observations in H\,{\sc i} emission using the Lovell and Arecibo telescopes 
(Kennedy et al. 1998; Smoker et al. 2001a) have shown that 
the IVC consists of several condensations of gas spread out over an area of more than 3 
square degrees, with structure existing down to the previous resolution limit of $\sim$ 3 arcmin. 
The brightest component is located towards M\,15 itself and has 
peak H\,{\sc i} column density at the Arecibo resolution of $\sim$ 
8$\times$10$^{19}$ cm$^{-2}$. In this paper, we study this part of the IVC 
at higher spatial resolution. 
The mass of this particular clump is $\sim$ 20 $D^{2}$ M$_{\odot}$, 
(where $D$ is the distance in kpc), thus for this particular object, in the absence of 
H$_{2}$, there is insufficient neutral gas 
to form an early-type star. Low-resolution absorption-line 
Ca\,{\sc ii} and Na\,{\sc i} spectroscopy (Lehner et al. 1999) towards cluster stars tentatively 
found cloud structure (or variations in the relative abundance) over scales as small as a few 
arcsec, with fibre-optic array mapping in the Na\,{\sc i} D absorption lines (Meyer \& Lauroesch 
1999) obtaining similar results with structure visible on scales of 
$\sim$ 4 arcsec (velocity resolution $\sim$ 16 km\,s$^{-1}$). Using empirical relationships 
between the sodium and hydrogen column densities, Meyer \& Lauroesch (1999) derived values of the 
H\,{\sc i} column density towards the cluster centre of $\approx$ 5$\times$10$^{20}$ cm$^{-2}$, 
some 6 times higher than that found using the Arecibo H\,{\sc i} data alone; the difference may be 
attributable to fine-scale cloud structure. Assuming spherical symmetry, a volume density 
of $\approx$ 1000 cm$^{-3}$ is implied by these latter results, similar to values obtained for gas 
in the local ISM (e.g. Faison et al. 1998, although see Lauroesch, Meyer \& Blades 2000). 
Such a high volume density and implied overpressure with respect to 
the ISM perhaps indicates that the assumption of spherical symmetry is invalid and that 
there may be some sheet like geometry in the IVC as has been postulated for low-velocity 
gas (Heiles 1997). 

In the current paper, we extend our studies of the M\,15 IVC to higher 
resolution and different wavelength regions in order to investigate 
three areas. Firstly, H\,{\sc i} synthesis 
mapping, WHAM H$\alpha$ and IRAS ISSA survey data retrieval towards the IVC 
were performed to see if the H\,{\sc i}, H$\alpha$ and 
infrared properties of the M\,15 IVC are compatible with either low 
velocity gas or HVCs in general, and whether there are any differences between 
the types of object, perhaps attributable to differences in the formation 
mechanism or current environment (for example, distance from the ionising field 
of the Galaxy). Secondly, wide-field medium-resolution H\,{\sc i} data were obtained in 
search of more IVC components, to trace how the gas kinetic temperature changes with 
sky position, and possibly determine the relative distance of cloud components from
the Galactic plane (c.f. Lehner 2000). Thirdly, longslit echelle Ca\,{\sc ii} 
observations were undertaken, using the centre of M\,15 as a background continuum source, 
in order to look for small-scale velocity and column density substructure within the 
IVC which could indicate the presence of cloudlets, collisions between which in certain 
IVCs may be responsible for star formation in the Galactic halo. 

Section \ref{observations} describes the observations and data reduction, 
Sect. \ref{results} gives the results, Sect. \ref{disc} contains the discussion and 
Sect. \ref{concl} presents a summary and the conclusions. 

\section{Observations and data reduction}
\label{observations}

\subsection{Westerbork aperture synthesis H\,{\sc i} observations}
\label{wsrtobs}

21-cm aperture synthesis H\,{\sc i} observations of the M\,15 IVC were obtained during 
two observing sessions, each of 12 hours, using the Westerbork Synthesis 
Radio Telescope (WSRT). The first in 19 December 1998 had a minimum antenna 
spacing of 32-m, the second in 17 April 1999 had a corresponding separation of 
72-m. The velocity resolution for all observations was 1.03 km\,s$^{-1}$. 

Standard methods within {\sc aips}\footnote{{\sc aips} is distributed by the 
National Radio Astronomy Observatory, U.S.A.} were used to reduce the visibility 
data. Reduction included amplitude calibration using 3C 286 and 3C 48 (assuming 
flux densities of 14.76 Jy and 15.98 Jy, respectively), phase calibration and 
flagging of bad data, and concatenation of the two UV datasets using {\sc dbcon}.  
The calibrated dataset was mapped with {\sc imagr} using both quasi-natural 
and uniform weightings, with the Arecibo map from Smoker et al. (2001a) being used 
to set the locations of the {\sc clean} boxes for each velocity channel interactively. The 
respective beamsizes of the final images were 
111$^{\prime\prime}\times$56$^{\prime\prime}$ 
and 81$^{\prime\prime}\times$14$^{\prime\prime}$ (approximately North--South by East--West), 
with the corresponding rms noises being 2.3 mJy beam$^{-1}$ and 1.3 mJy beam$^{-1}$ 
for the naturally and uniformly-weighted data. 

As the WSRT maps suffer from missing short-spacing information, it was decided 
to create `total power' channel maps by combining the current observations with the 
single dish Arecibo data. For this we used {\sc miriad} (Sault, Teuben \& Wright 1995). 
The two datasets were first regridded so that they had the same central coordinate and channel width. 
Following this, we multiplied the Arecibo data by the WSRT single-dish beam, converted to 
flux density units, and combined the WSRT and weighted Arecibo maps using 
{\sc immerge}. Combination was performed by specifying an annulus of between 
50--150-m in the Fourier domain where the high and low resolution images
were made to agree. Within this annulus it was found that, where the 
H\,{\sc i} was strong, the flux density scales of the two datasets were 
the same. Hence the `calibration factor' used when combining the two datasets was
set to unity. The resulting combined channel maps were divided by the WSRT single 
dish beam to produce the final combined images. These are of the 
resolution of the WSRT data (111$^{\prime\prime}\times$56$^{\prime\prime}$), but with 
the correct zero-spacing flux. Finally, moments analysis and Gaussian fitting were  
performed on the WSRT channel maps and combined cube from +50 to +90 km\,s$^{-1}$ 
in order to determine the H\,{\sc i} column densities, velocities and velocity widths 
of the IVC. H\,{\sc i} column densities were derived using the relationship 
N$_{HI}$=1.823$\times$10$^{18} \int$T$_{\rm B} dv$ (cm$^{-2}$), where $v$ 
is the velocity in km\,s$^{-1}$. 

\subsection{Lovell Telescope multibeam H\,{\sc i} observations}

\begin{figure}
\epsffile{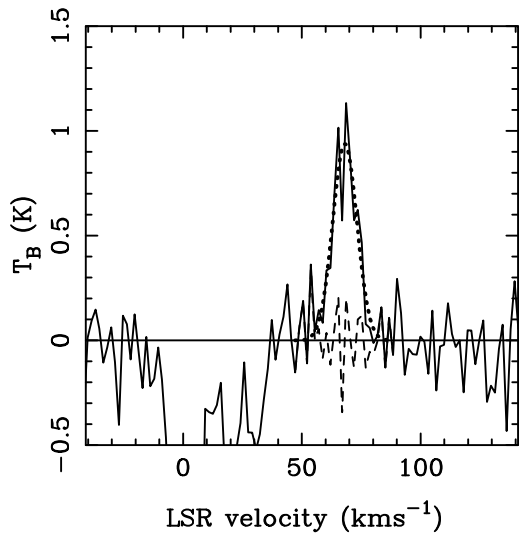}
\caption{Lovell telescope multibeam H\,{\sc i} spectrum towards 
RA=21$^{h}$26$^{m}$47$^{s}$, Dec=+09$^{\circ}$16$^{\prime}$23$^{\prime\prime}$ 
(J2000). The dotted line shows the results of a single-component Gaussian fit. The brightness 
temperature calibration is accurate to $\sim$ 40 per cent with the negative 
features present near to $\sim$ 0 km\,s$^{-1}$ being an artefact of the baseline 
removal procedure. The dashed line is the (data$-$single Gaussian fit) residual. 
}
\label{mbspec}
\end{figure}

Lovell telescope multibeam 21-cm H\,{\sc i} observations covering a $\sim$ 4.9$\times$9.3 
degree RA$-$Dec region, centred upon 
RA=21$^{h}$22$^{m}$47$^{s}$, Dec=+10$^{\circ}$00$^{\prime}$29$^{\prime\prime}$ 
(J2000) were taken during 2 June 2000. The resolution of these data is 12 arcmin spatially 
and 3.0 km\,s$^{-1}$ in velocity. The total integration time was 4 hours and the RMS noise 
per channel was $\sim$ 0.1 K. The observed field is not quite fully sampled, with the 
spacing between the beams being 10 arcmin, compared with the spatial resolution 
of 12 arcmin. During observing, data were calibrated in terms of antenna temperature 
by a frequently-fired noise diode. Off-line, bootstrapping of the flux scale to some previous 
Lovell telescope H\,{\sc i} observations taken during 1996 enabled calibration in terms 
of brightness temperature (T$_{\rm B}$). Data reduction procedures will be discussed in more 
detail in a future paper. For now, we simply note that most of the process is automated, 
the bandpass removal being performed within {\sc aips$^{++}$} using {\sc livedata} and 
the data being gridded using {\sc gridzilla} (Barnes 2001). As the bandpass is simply a running 
average of the observed spectra in a window, incorrect removal is performed around the 
low-velocity gas where there is no `empty' region of the sky. However, at 
the velocity of the IVC (v $\sim$ +70 km\,s$^{-1}$) the baseline subtraction is adequate. 
An example of a spectrum after automated bandpass removal is shown in Fig. \ref{mbspec}. 

After gridding, the data were imported into {\sc aips} whence a column density map was 
made of the entire field between +50 and +90 km\,s$^{-1}$ which is presented 
in section \ref{ltmbresults}. 

\subsection{William Herschel Telescope longslit UES observations}

WHT Utrecht Echelle Spectrograph (UES) observations towards the centre of the M\,15 
globular cluster were taken during bright time on 19--21 July 2000. Twin EEV CCDs were used. 
In combination with the E79 grating, this gave near-complete wavelength coverage from 
3800--6300 \AA, with a slitwidth of 0.85$^{\prime\prime}$ providing an instrumental FWHM 
resolution of $\sim$ 6 km\,s$^{-1}$. On the first night of the run, a slit height of 
16 arcseconds was used; this has the advantage that the full wavelength coverage is obtained 
with no overlapping orders on either of the CCDs. On the second and third night's observations, 
in order to maximise the amount of sky available for the calcium lines at 3933.66\AA, a slit 
of height 22 arcseconds was used which leads to overlapping orders in the red CCD. ThAr arcs 
were taken two or three times a night to act as the wavelength calibration with Tungsten 
and sky flats also being obtained. The echelle proved to be remarkably stable with the 
centres of the arc lines varying by less than 1.5 pixels throughout the whole of the run. 

The original intention of the observations was to obtain 7 parallel cuts through the IVC, 
separated by 0.7 arcsec, and use these to make up an RA--Dec rectangle of size 
$\sim$ $5^{\prime\prime} \times 20^{\prime\prime}$ within which cloudlet sizes could be 
obtained. Unfortunately, due to the presence of Sahara 
dust and non-ideal seeing conditions ($1.5^{\prime\prime}$--$2.0^{\prime\prime}$), integration 
times were somewhat longer than had been hoped for and in the end only two longslit positions 
were obtained, offset by a position angle of 85 degrees. Additionally, on the second night we 
obtained 1200-s of integration towards a blank sky region with a 0.85$^{\prime\prime}$ 
slit and on the third night a total of 5$\times$400-s, interleaved with the object exposures 
and using a slit of width 4.0$^{\prime\prime}$, corresponding to $\sim$9500-s of 
integration with a slit of 0.85$^{\prime\prime}$ used for the IVC observations. 

Data reduction was performed using standard methods within {\sc iraf}{\it \footnote{{\sc iraf} 
is distributed by the National Optical Astronomy Observatories, U.S.A.}}. The aims were  
firstly to obtain a spectrum of the inner part of the globular cluster over the whole
wavelength region, and more importantly, produce longslit spectra of the Ca\,{\sc ii} 
lines alone for the two longslit positions taken. Data reduction included debiasing, 
and use of the sky flatfields to get rid of the vignetting that was clearly visible on many of the 
orders. Cosmic ray hits near the Ca\,{\sc ii} line were removed within {\sc figaro} 
(Shortridge et al. 1999) using {\sc clean}. Because of the small pixel-to-pixel variation 
and pixel size ($\sim$ 0.2$^{\prime\prime}$), no flatfielding to remove the pixel-to-pixel 
variation was performed on the data. The images were rotated in order to make the 
cross-dispersion axis occur along the image rows, after which sky subtraction was performed. 
As expected, this proved to be challenging. 

\subsubsection{Sky subtraction}

\begin{figure}
\epsfxsize=8 truecm
\epsffile{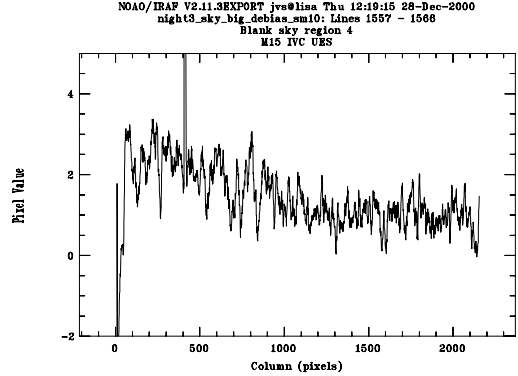}
\caption{Cut across the dispersion axis at $\sim$ 3937\AA\, of a blank sky region, 
equivalent to 5400-s of object observing time on night 3. The `dips' in the spectrum 
correspond to the interorder gaps.
}
\label{sky1}
\end{figure}

Two methods were used to estimate the sky value and the results compared. The first  
used the data from the third nights observation where the blank sky exposures were 
interleaved with the object exposures. For these data, we scaled the high signal-to-noise 
combined sky flat (taken at twilight with the 0.85 arcsec slit) to produce an image `scaled\_sky'
and then smoothed this in wavelength to a resolution equivalent to a 4 arcsecond slit. 
The scaling factor was chosen so that the resulting image was the same as the blank sky 
image taken with the 4 arcsecond slit for an equivalent of 5400-s of integration. The 
image `scaled\_sky' was then subtracted from each 5400-s of the object data.
This method has the advantage that in the blank sky images there 
are clearly no contaminating counts from the object, 
but is susceptible to changes in the observing conditions and requires careful removal of 
the bias strip in both sky and object frames. Fig. \ref{sky1} shows a cut across the 
dispersion axis for the sky exposures on night 3. As can be seen, there are few sky counts, 
although obviously in the outer parts of the slit there are few object counts too, so errors 
in the equivalent width increase rapidly in the extracted spectra for 
this region. For an equivalent 5400-s of integration 
time, at $\lambda \sim$ 3937\AA, there were $\sim$ 3 ADU sky counts and somewhat less at 
$\lambda$=3933.66\AA, due to the `sky' effectively being a solar (G-type) spectrum which has 
strong Ca\,{\sc ii} absorption lines. 

\begin{figure}
\epsffile{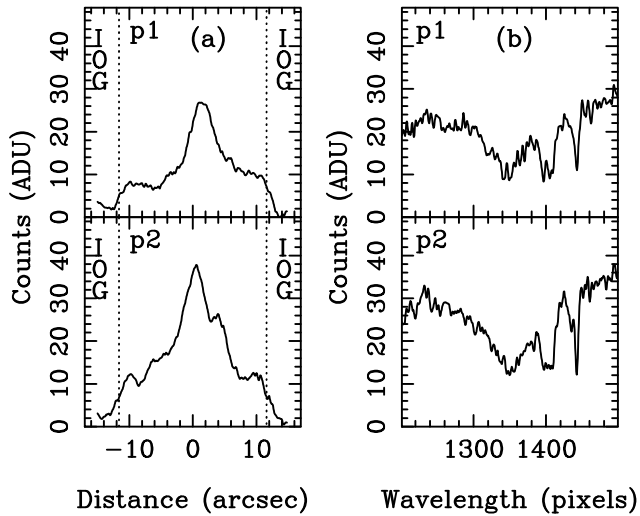}
\caption{(a) Cut across the dispersion axis at $\sim$ 3937\AA\ for the two slit 
positions with 5400-s of integration, showing the difficulty in determining the sky. 
IOG refers to the interorder gap between the echelle orders (b) Cut along the dispersion 
axis before normalisation. 
}
\label{whtcross}
\end{figure}

The method of sky subtraction that we had envisaged performing {\em a priori} involved removal 
of sky using parts of the longslit that appeared to be free of emission from the globular cluster. 
The problem with this method is illustrated by Fig. \ref{whtcross}(a), which shows cuts across 
the dispersion axis for 5400-s of data taken on the third night at each of the two slit positions. 
As is apparent from the Figures, the `sky' level at the same position on the order 
at the two different slit positions is different by $\sim$ 3--4 ADU per 5400-s of integration. 
During night 3, at $\lambda$ $\sim$ 3937\AA, position p1 had $\sim$ 8 ADU as its `sky' value, 
with the corresponding value for p2 being $\sim$ 12 ADU. These clearly are much larger than the 
values obtained by exposing on the night sky and combined with the fact that p1 and p2 are different  
indicates that we are still obtaining counts from the cluster in the outer parts of the slit. 
This is confirmed by performing an extraction using such counts; in some parts of the spectrum 
the IVC absorption becomes negative which is unphysical. 

Because of the lack of object-free emission on the slit, we decided to remove the sky by subtracting 
the smoothed version of the sky images from our object data. This method relies upon the sky 
conditions being similar throughout the three nights of observing. This was checked by taking 
cuts across three places on the dispersion axis of each 5400-s worth of data and checking for 
variability. At $\lambda \sim$ 3937 \AA, the values varied by only 2 ADU per 5400-s between nights 
1 and 3 for the two slit positions. This variability, combined with the error in removing the bias, 
gives us an error in the final sky value of $\sim$ 4 ADU at 3937\AA, compared with a {\em peak} 
continuum value at this point of $\sim$ 30 ADU. 
  
\subsubsection{Extraction and Analysis}
\label{lsana}

The sky-subtracted spectra were extracted within {\sc figaro} by summing over each 
10 columns, corresponding to the worst seeing of $\sim$ 2.0 arcseconds. Wavelength 
calibration was then undertaken, taking into account the shift in the $\lambda$ 
scale along the length of the slit. Finally, normalisation and profile fitting 
were performed within {\sc dipso} (Howarth et al. 1996). The final spectra have 
spatial resolution of $\sim$ 2 arcseconds and velocity resolution of 
$\sim$ 7 km\,s$^{-1}$.

Data were analysed using standard profile fitting methods within 
{\sc dipso} using the {\sc elf} and {\sc is} programs. Before fitting, the 
the stellar line was removed by fitting the profile by eye to the lower-wavelength 
part of the spectrum up to the peak stellar absorption unaffected by LV and IV gas, 
and creating a mirror of this spectrum for the higher wavelength data. The derived 
stellar spectrum has a stark-like line profile, typical of stellar Ca\,{\sc ii} lines,  
and was subtracted from the whole spectrum to leave the interstellar components only. 
After stellar-line removal, the {\sc elf} routines were used to 
fit Gaussian profiles to input spectra and hence provide the equivalent width, 
peak absorption and full width half maximum velocity values for the low 
velocity gas and IVC components. These results were compared to the theoretical 
absorption profiles computed by the {\sc is} suite of programs within {\sc dipso}. 
These theoretical profiles are derived from the observed $b$-values (corrected for 
instrumental broadening), atomic data for the Ca\,{\sc ii} lines taken from Morton 
(1991) and initial guesses for the Ca\,{\sc ii} column density. Input parameters were varied 
until a good fit was produced to match the profiles obtained using {\sc elf}. This 
method is only appropriate in the regime where the lines are not saturated and
are resolved in velocity; both caveats apply for a number of positions in 
the current dataset, although the presence of narrow, unresolved components cannot 
be ruled out. If present, these would cause our estimated column densities to 
be too low. The final {\sc elf} and {\sc is} fits provide the Ca\,{\sc ii} number densities  
and values for the FWHM velocities at each of the positions along the slit. 

\subsection{Wisconsin H$\alpha$ Mapper observations}

Data were retrieved from the Wisconsin H$\alpha$ Mapper (WHAM) 
survey in the range +06$^{\circ}$ $<$ Dec. $<$ +15$^{\circ}$ and 20$^{h}$20$^{m}$ $<$ RA 
$<$ 21$^{h}$38$^{m}$ (J2000). These data are of resolution 1 degree spatially and 
$\sim$ 12 km\,s$^{-1}$ in velocity, with a velocity coverage at this sky position from 
$\sim$ $-$120 to +90 km\,s$^{-1}$ in the LSR. Data were reduced using standard methods, which 
included conversion into the LSR and removal of the geocoronal H$\alpha$ line that appears 
at velocities in the range $\sim$+30 to +40 km\,s$^{-1}$ (Haffner 1999; Haffner et al. 2001b). 
Due to the removal of this line, the noise in this part of 
the spectrum is slightly enhanced compared with the typical RMS 
noise value (measured for the current spectra) of $\sim$1$-$3 mR 
(km\,s$^{-1}$)$^{-1}$. These values are typical for the WHAM survey
as a whole. Finally, we note if the IVC H$\alpha$ flux originates from above the Galactic plane, 
then a correction for Galactic extinction of 1.25 would be necessary, calculated from a 
E($B-V$)=0.10 observed towards M\,15 by Durrell \& Harris (1993), combined with 
the extinction law of Cardelli, Clayton \& Mathis (1989). We have not applied this 
correction to our data so the IVC H$\alpha$ fluxes may be $\sim$ 25 per cent too low. 

\subsection{IRAS ISSA data retrieval}

Both the 60 and 100 micron images of a field of size 0.675$^{\circ} \times$0.625$^{\circ}$ 
were extracted from the on-line versions of the IRAS Sky Survey Atlas (ISSA). These 
images are of resolution $\sim$ 5 arcmin and have most of their zodiacal emission removed 
although they have an arbitrary zero-point. 

\section{Results}
\label{results}

\begin{figure}
\epsffile{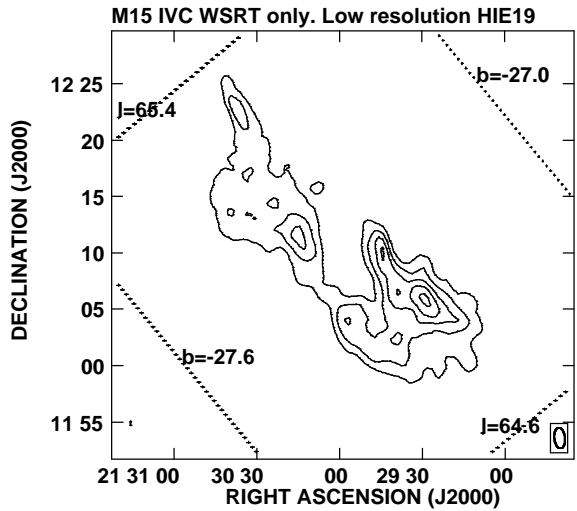}
\caption{WSRT H\,{\sc i} column density map of the M\,15 IVC 
integrated between +50 and +90 km\,s$^{-1}$ 
and at a resolution of 111$^{\prime\prime}\times$56$^{\prime\prime}$ 
($\sim$ NS by EW). Contour levels are at 
N$_{HI}$=(2,4,6,8,10)$\times$10$^{19}$ cm$^{-2}$.
}
\label{wsrtlowhisur}
\end{figure}

\begin{figure}
\epsffile{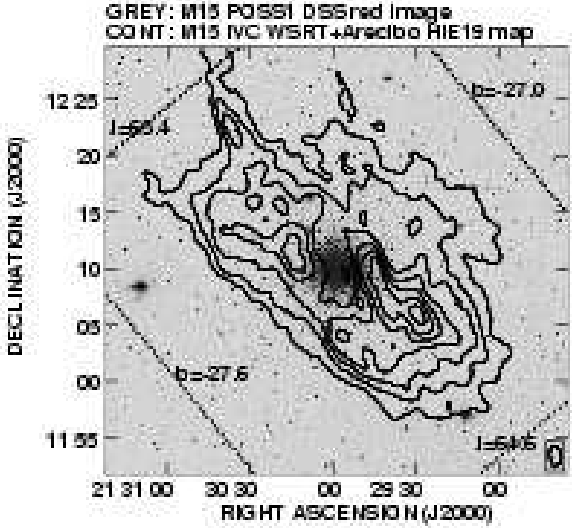}
\caption{WSRT plus Arecibo combined H\,{\sc i} column density map of the M\,15 
IVC integrated between +50 and +90 km\,s$^{-1}$ and at a resolution of 
111$^{\prime\prime}\times$56$^{\prime\prime}$, overlaid on the 
red POSS-1 digital sky survey image towards M\,15. Velocities are in the LSR with 
contour levels being at N$_{HI}$=(2,4,6,8,10,12,14)$\times$10$^{19}$ cm$^{-2}$.
Lines of constant Galactic longitude and latitude are also plotted on the 
figure.
} 
\label{combhisur}
\end{figure}

\subsection{WSRT aperture synthesis H\,{\sc i} results}
\label{wsrtresults}

\begin{figure*}
\epsffile{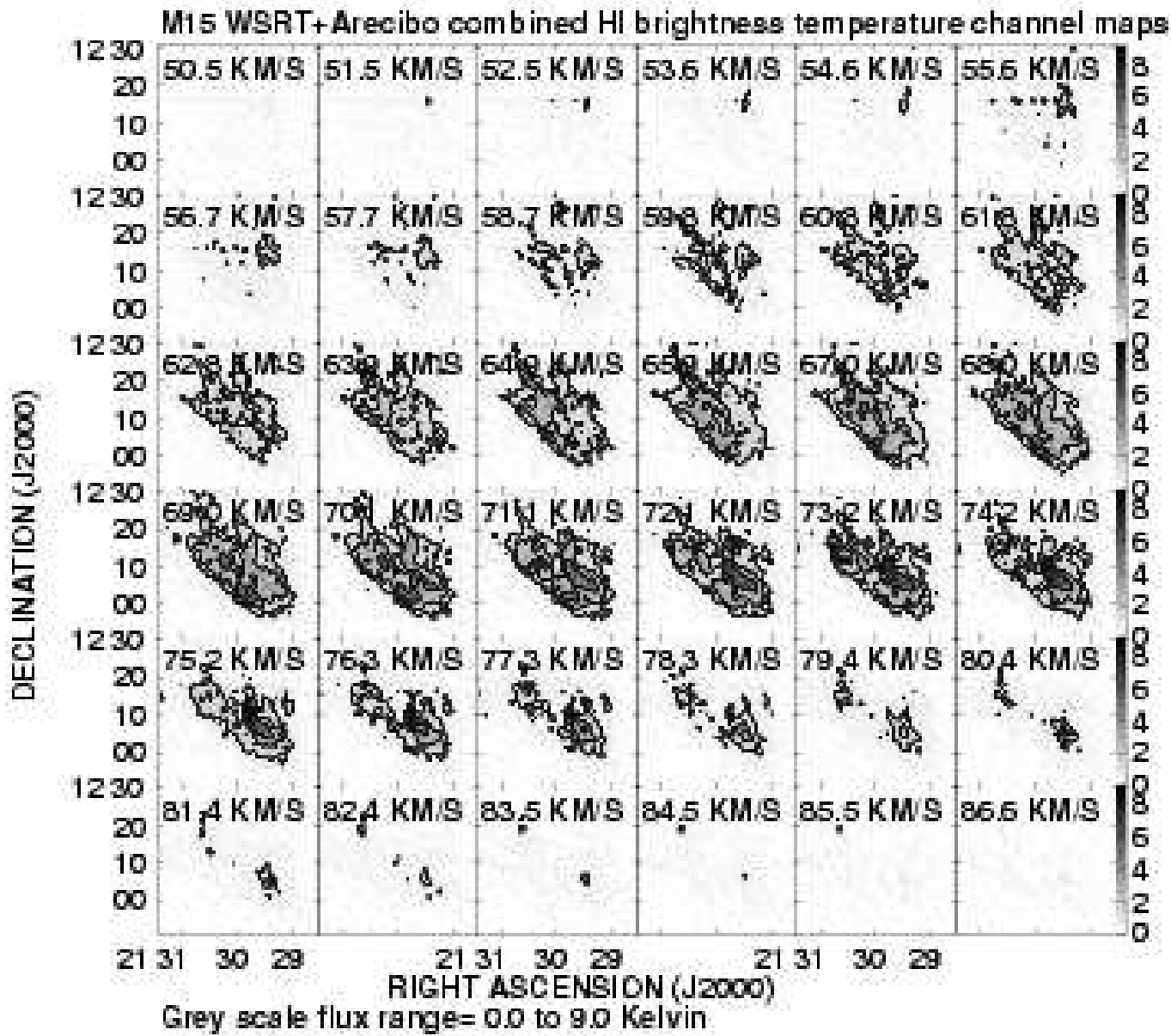}
\caption{WSRT plus Arecibo combined H\,{\sc i} brightness temperature channel maps
towards the M\,15 IVC. Velocities are in the LSR and the velocity resolution is 
1.03 km\,s$^{-1}$. Contour levels are at T$_{B}$=(--1,1,2,4,8) Kelvin. The spatial resolution is 
111$^{\prime\prime}\times$54$^{\prime\prime}$ ($\sim$ NS by EW).
} 
\label{combchantb}
\end{figure*}

Fig. \ref{wsrtlowhisur} shows the H\,{\sc i} column density map of the IVC for the 
low-resolution WSRT data alone, with Fig. \ref{combhisur} showing the corresponding map 
of the combined (WSRT plus Arecibo) image overlaid on the digitised Palomar 
Observatory Sky Atlas (POSS-I) red image regridded to a 4 arcsec pixel size. The major 
axis of this part of the IVC lies parallel with the Galactic plane. 
The H\,{\sc i} channel maps of the combined WSRT plus Arecibo dataset in brightness 
temperature (T$_{\rm B}$) are shown in Fig. \ref {combchantb}, where flux density per 
beam is related 
to T$_{\rm B}$ by S$_{\rm mJy/beam}$=0.65$\Omega_{\rm as}$T$_{\rm B}$/$\lambda_{cm}^{2}$;  
here $\Omega_{\rm as}$ is the beam area in arcsec$^{2}$ and 
$\lambda_{cm}^{2}$ is the observed wavelength in cm (Braun \& Burton 2000). 
Immediately obvious from each of these figures is the fact that the H\,{\sc i} is 
clumpy in nature, as is seen in other intermediate and high velocity 
clouds observed at a similar resolution (e.g. Wei$\beta$ et al. 1999; Braun \& Burton 2000).
Variations in the column density of a factor of $\sim$ 4 on scales of $\sim$ 5 arcmin  
are observed, corresponding to scales of $\sim$ 1.5 $D$ pc, where $D$ is the 
the IVC distance in kpc. The peak brightness temperature in the combined map (of 
resolution 111$^{\prime\prime}\times$56$^{\prime\prime}$), is $\sim$ 8 K, rising to 
$\sim$ 14 K for the highest resolution WSRT map with beamsize 
81$^{\prime\prime}\times$14$^{\prime\prime}$. These values compare with 
peak values of T$_{B}$ for HVCs of $\sim$ 25 K towards Complex A at 2 arcmin resolution 
(Schwarz, Sullivan \& Hulsbosch 1976), $\sim$ 34 K observed towards Complex M at a 
resolution of 1 arcmin (Wakker \& Schwarz 1991), and $\sim$ 75 K observed 
to the compact high velocity cloud CHVC 125+41--207 (Braun \& Burton 2000). 
Similarly the IVC observed by Wei$\beta$ et al. (1999) has peak T$_{B}$ 
exceeding 7 K, at a resolution of 1 arcmin. 

The peak H\,{\sc i} column density in the combined image is 
$\sim$ 1.5$\times$10$^{20}$ cm$^{-2}$. If we assume that the cloud is spherically 
symmetric or filamentary (Stoppelenburg, Schwarz \& van Woerden 1998), and using a 
cloud size full width half maximum (c.f. Wakker \& Schwarz 1991) of $\sim$ 7 arcmin or 
2 $D$pc for the brightest point, we obtain 
a peak volume density of $\sim$ 24$D^{-1}$ cm$^{-3}$. At the assumed distance, this value is 
a lower limit as it is likely that there will be more structure on smaller scales, 
as indicated by the spectra of Meyer \& Lauroesch (1999). 

\begin{figure}
\epsffile{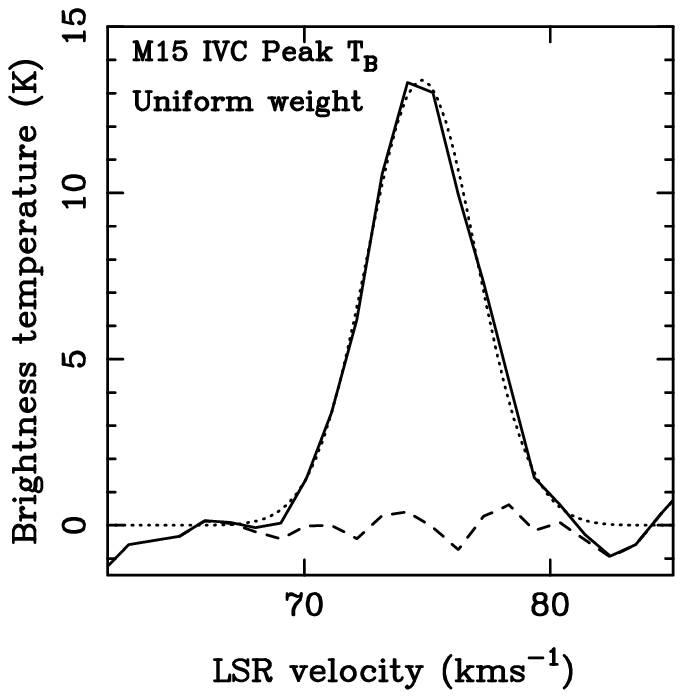}
\caption{WSRT uniformly-weighted H\,{\sc i} brightness temperature 
profile of the M\,15 IVC at the position of peak T$_{\rm B}$. The beamsize is 
81$^{\prime\prime}\times$14$^{\prime\prime}$. The dotted line depicts 
the Gaussian fit to the spectrum and has FWHM=5.2$\pm$0.2 km\,s$^{-1}$. 
The dashed line shows the (data$-$single Gaussian fit) residual.
}
\label{unitbint}
\end{figure}

\begin{figure}
\epsfxsize=8.4 truecm
\epsffile{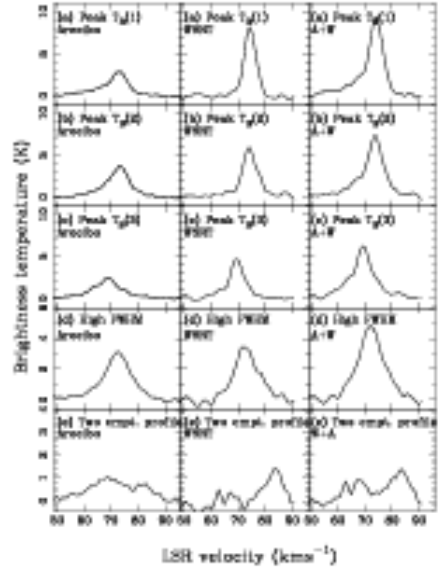}
\caption{WSRT plus Arecibo combined H\,{\sc i} brightness temperature 
profiles towards selected regions of the M\,15 IVC. (a) First T${_B}$ peak  
(RA=21$^{h}$29$^{m}$44$^{s}$, Dec.=+12$^{\circ}$09$^{\prime}$42$^{\prime\prime}$). 
(b) Second T${_B}$ peak (RA=21$^{h}$29$^{m}$36$^{s}$, 
Dec.=+12$^{\circ}$06$^{\prime}$22$^{\prime\prime}$). (c) Third T${_B}$ peak 
(RA=21$^{h}$30$^{m}$13$^{s}$, Dec.=+12$^{\circ}$10$^{\prime}$42$^{\prime\prime}$). 
(d) High FWHM region (RA=21$^{h}$29$^{m}$21$^{s}$, 
Dec.=+12$^{\circ}$04$^{\prime}$22$^{\prime\prime}$).  
(e) Two component profile (RA=21$^{h}$30$^{m}$38$^{s}$, 
Dec.=+12$^{\circ}$19$^{\prime}$22$^{\prime\prime}$). The equinox is J2000.
}
\label{combtbint}
\end{figure}

Fig. \ref{unitbint} shows the brightness temperature (T$_{\rm B}$) profile at 
the position of peak temperature in the uniformly-weighted data, with 
Fig. \ref{combtbint} depicting the combined temperature profiles at a 
number of positions in the cloud. The full width half maximum (FWHM) varies 
from 5 km\,s$^{-1}$ at the position of the three peaks in brightness temperature
(Fig. \ref{unitbint} and Fig. \ref{combtbint}(a)--(c)), to more than 12 km\,s$^{-1}$ at 
other locations in the IVC (Fig. \ref{combtbint}(d)). Some parts of the cloud (cf Fig. 
\ref{combtbint}(b)) are well-fitted by a Gaussian, whereas others, such as Figs. 
\ref{combtbint}(e), appear to be made up of two narrowish components. Although the 
signal-to-noise in this latter part of the cloud is low, both the WSRT and Arecibo datasets 
show evidence for a two-component velocity substructure, perhaps indicative of 
H\,{\sc i} cloudlets or of overlapping clouds in the line of sight. 

\begin{figure}
\epsffile{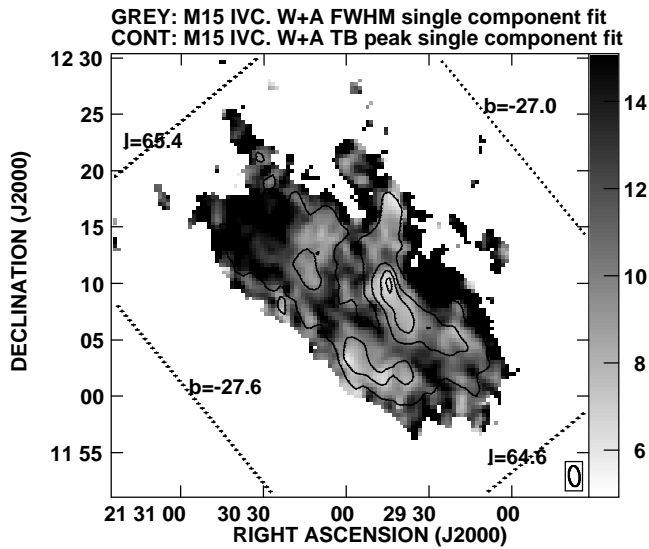}
\caption{WSRT plus Arecibo H\,{\sc i} combined image. Greyscale; full width half maxima 
velocity widths obtained using single-component Gaussian fitting. Contours; peak values of 
brightness temperature of the same single-component fits with levels at (2,4,6,8) Kelvin. 
}
\label{fwhmtb}
\end{figure}

Fig. \ref{fwhmtb} displays the results of single-component Gaussian fitting to the IVC 
using {\sc xgauss} and shows how the smaller values of FWHM (and hence implied kinetic 
temperature) tend to occur where the H\,{\sc i} is brightest in the three cores. 
At these positions, in the absence of turbulence, 
the kinematic temperature is of the order of 500 K. The M\,15 IVC does not show any regular 
velocity gradients across its field as has been seen in many other HVCs and IVCs 
(e.g. Wei$\beta$ et al. 1999; Br\"uns et al. 2000). 

We finally note that an interesting feature of the H\,{\sc i} surface density map 
is that the IVC is approximately centred upon the globular cluster, with the 
regions of lowest column density being located towards the globular cluster centre. 
It is tempting to suggest some mechanism whereby the IVC and globular cluster are
associated, for example by capture of the IVC by M\,15, 
with the gas in the centre being ionised by the cluster itself. However, the
existence of other IVC components in the nearby field, combined with the fact that 
M\,15 is located at a radial velocity of $\sim$ --100 km\,s$^{-1}$
(Harris 1996), compared with the the IVC at $\sim$ +70 km\,s$^{-1}$, makes it likely that
the two objects are simply line-of-sight companions. 

\subsection{Lovell telescope multibeam H\,{\sc i} results}
\label{ltmbresults}

\begin{figure}
\begin{center}
\epsffile{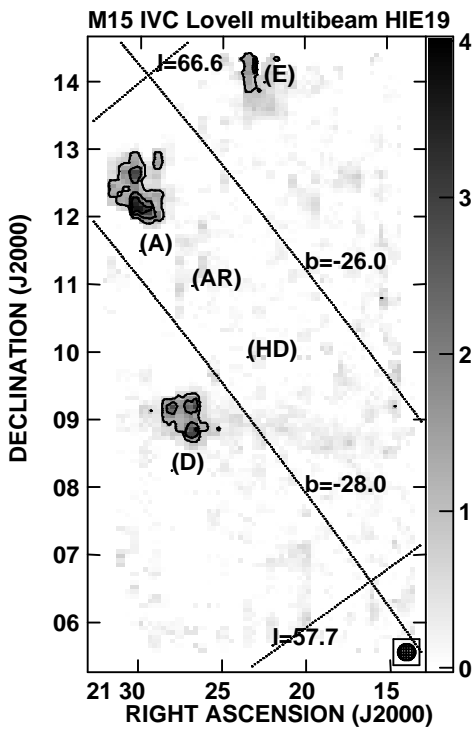}
\end{center}
\caption{Lovell telescope multibeam H\,{\sc i} column density map of 
resolution 12 arcmin in the general direction of the M\,15 globular cluster integrated 
from +50 to +90 km\,s$^{-1}$ in the LSR. Contour levels are at 
(1,2,3)$\times$10$^{19}$ cm$^{-2}$, with greyscale levels from 0--4$\times$10$^{19}$ 
cm$^{-2}$. For the meaning of the labels see Sect. \ref{ltmbresults}. 
}
\label{mbhicol}
\end{figure}

The H\,{\sc i} column density map of a 4.9$\times$9.3 degree field, with 
spatial resolution 12 arcmin, centred to the South-West of M\,15, integrated 
from +50 to +90 km\,s$^{-1}$, is shown in Fig. \ref{mbhicol}. 

No new firm H\,{\sc i} detections were obtained that are not 
already given in the Leiden/Dwingeloo survey (Hartmann \& Burton 1997), 
Kennedy et al. (1998) or Smoker et al. (2001a). However, the extent of an IVC component 
around RA=21$^{h}$27$^{m}$, Dec=+09$^{\circ}$10$^{\prime}$ (J2000; feature `D') was determined 
to be $\sim$ 0.7 degrees in RA--Dec at full width half power in the N$_{HI}$ map. An example 
H\,{\sc i} spectrum in this area is shown in Fig. \ref{mbspec}. 
At this position, a single-component Gaussian fit within {\sc dipso} gives a velocity centroid of 
68$\pm$0.5 km\,s$^{-1}$, FWHM of 12.0$\pm$1.2 km\,s$^{-1}$ and peak T$_{B}$ of 
$\sim$ 1 K, similar properties to the main complex studied in this paper. The peak IVC 
H\,{\sc i} column density towards this component (`D') is $\sim$ 3$\times$10$^{19}$ cm$^{-2}$. 

Also plotted on Fig. \ref{mbhicol} are `A', `AR' and `HD'. `A' corresponds to the main clump 
observed by Kennedy et al. (1998) which has FWHM $\sim$ 12--15 km\,s$^{-1}$ at 12 arcmin 
resolution. The peak IVC column H\,{\sc i} density derived using the current observations towards component 
`A' of 3.7$\times$10$^{19}$ cm$^{-2}$ is close to the value of 3.9$\times$10$^{19}$ cm$^{-2}$ 
observed by Kennedy et al. (1998) with the same telescope and gives us faith in our calibration. 
`AR' refers to previous IVC detections using the Arecibo telescope which have FWHM=12 km\,s$^{-1}$ and 
$v_{LSR}$=+61 km\,s$^{-1}$ at a resolution of 3 arcmin. `HD' is the position of the halo star HD 203664 (of 
distance 3 kpc) in which two strong interstellar IVC Ca\,{\sc ii} K absorption components are seen with 
FWHM=2.8 and 3.2 km\,s$^{-1}$ and v$\sim$ +80 and +75 km\,s$^{-1}$ respectively 
(Ryans, Sembach \& Keenan 1996). As this is an absorption-line measurement towards a star, 
the resolution is sub-arcsecond. Finally, we note that there is a hint of emission at 
RA=21$^{h}$23$^{m}$04$^{s}$, Dec=+14$^{\circ}$12$^{\prime}$42$^{\prime\prime}$ (feature `E' 
on Fig. \ref{mbhicol}) although this is very close to the noise and may be a spurious detection. 

\subsection{William Herschel Telescope longslit UES results}

\begin{figure}
\begin{center}
\epsffile{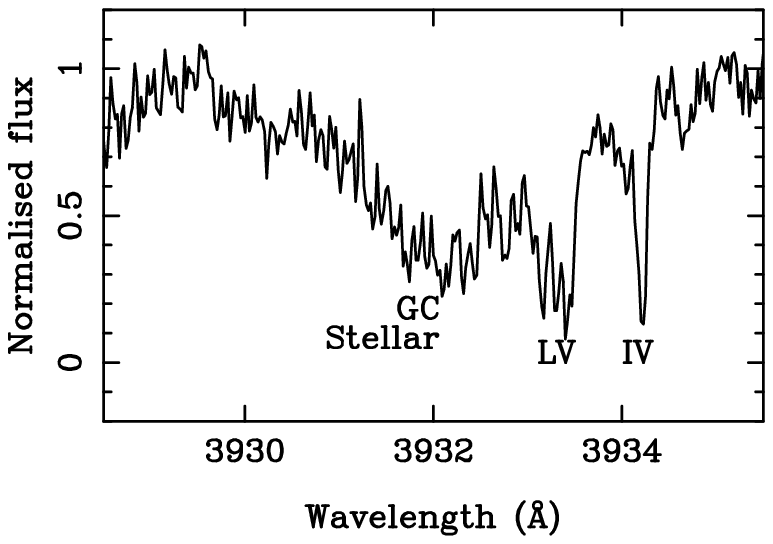}
\end{center}
\caption{Ca\,{\sc ii} K central 3$^{\prime\prime}$ towards M\,15. 
}
\label{caiicentre}
\end{figure}

Equivalent widths of some strong stellar lines from the inner 3 arcseconds of the 
first slit position are shown in Table \ref{ewstar}. As the core of the cluster is 
unresolved from the ground, the spectrum obtained is composite. Fig. 
\ref{caiicentre} shows the Ca\,{\sc ii} K spectra at this position.
The equivalent widths of the LV and IV interstellar components at this position are 
$\sim$ 0.3 \AA \, and $\sim$ 0.08 \AA \, respectively. The strength of the LV component 
multiplied by sin$(b)$ of $\sim$ 135 m\AA compares with the canonical value of 
$\sim$ 110 m\AA \, integrated EW of the  Ca\,{\sc ii} K perpendicular 
to the Galactic plane (Bowen 1991).

\begin{table}
\caption{Equivalent widths of some strong lines towards the 
centre of M\,15. Where there are two values present, these refer to results taken on nights 1 and 2 
respectively. Stellar types corresponding to the EWs for that particular 
line are taken from Jaschek \& Jaschek (1987) (JJ87).}
\centering
\label{ewstar}
\begin{tabular}{cccc}
Species         &     $\lambda_{\rm rest}$    &  EW (\AA)  & JJ87 Type \\
                &                         &                &           \\
Ca\,{\sc ii} K  &    3933.66              &  1.6,  1.5     & A2        \\
Fe\,{\sc i}     &    4045.82              & 0.16, 0.17     & A3        \\
Ca\,{\sc i}     &    4226                 & 0.25, 0.28     & F2        \\
\end{tabular}			    
\end{table}			   
\normalsize

\begin{figure}
\begin{center}
\epsffile{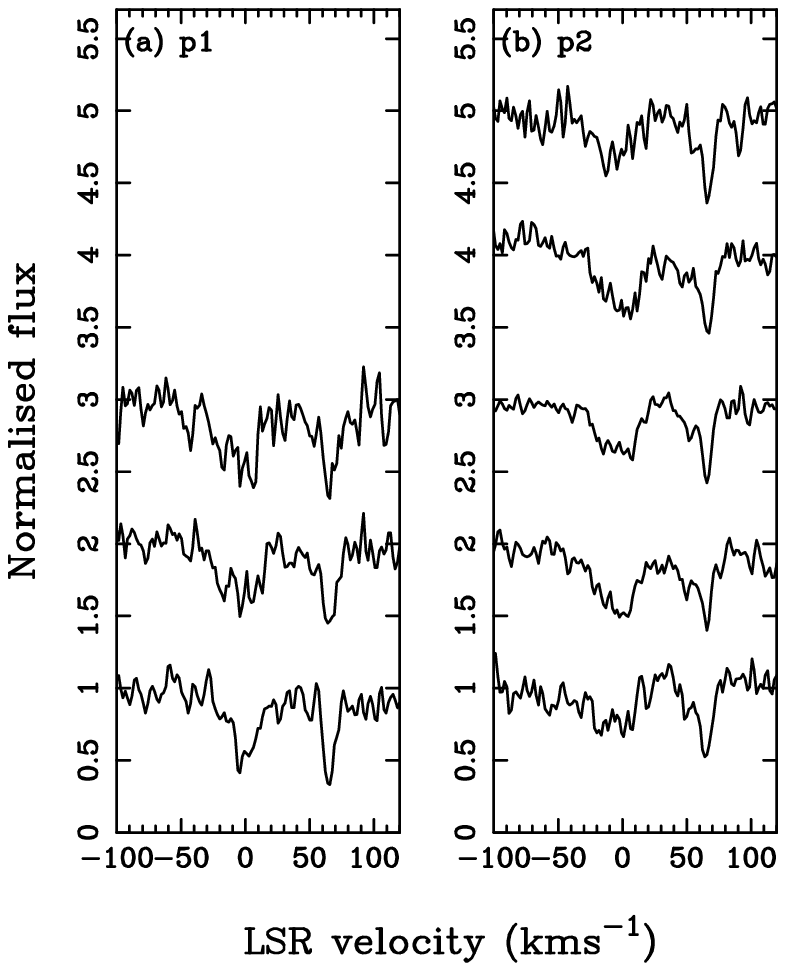}
\end{center}
\caption{Normalised spectra each 2 arcseconds along the slit. Spectra are 
offset in units of 1.0 for clarity. (a) Ca\,{\sc ii} K spectra at slit position 1. 
(b) Ca\,{\sc ii} K spectra at slit position 2. 
}
\label{caii_wht}
\end{figure}

We now turn to the interstellar Ca\,{\sc ii} observations. Fig. \ref{caii_wht} shows the 
extracted spectra at three and five positions along the dispersion axis for slit positions 
1 and 2 respectively, where the signal-to-noise in the continuum exceeds $\sim$ 10. 
The positions shown are separated by 2 arcseconds which corresponds to the worst seeing 
during the run, and is also the spatial resolution to which the data were smoothed when 
the extraction was performed. 

\begin{figure}
\begin{center}
\epsffile{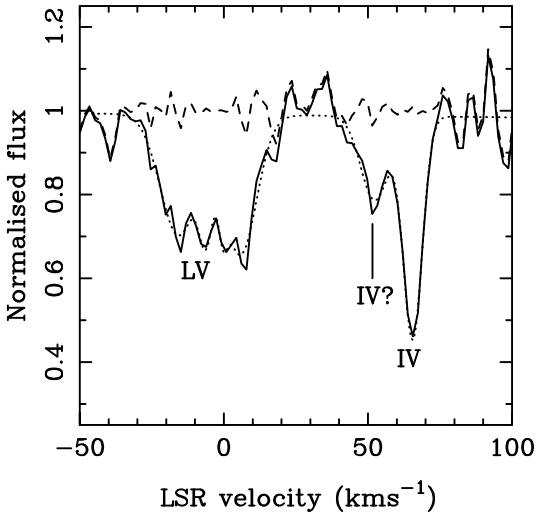}
\end{center}
\caption{Solid line: Ca\,{\sc ii} K LV and IV spectrum. Dotted line; {\sc elf} 
multicomponent fit. Dashed line; (data$-$model) residual plus 1.0. A possible 
second intermediate velocity component is marked by IV?.
}
\label{caii_wht_fit}
\end{figure}

\begin{figure}
\begin{center}
\epsffile{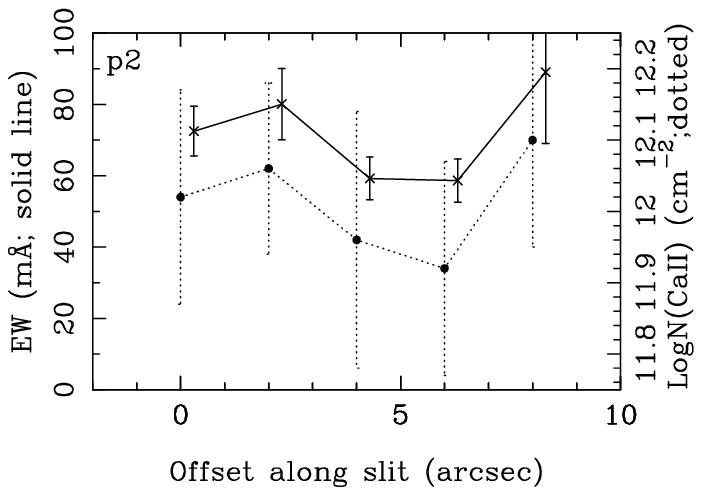}
\end{center}
\caption{Ca\,{\sc ii} fit results along the second slit position. 
}
\label{elf_fits}
\end{figure}

The prime aim of these longslit observations was to determine whether there is 
velocity substructure within the IVC. As can be seen from a few of the sightlines,
there is {\em tentative} evidence for such structure, with two cloudlets being 
present at LSR velocities of $\sim$ +52 km\,s$^{-1}$ and $\sim$ +66 km\,s$^{-1}$ 
(Fig. \ref{caii_wht_fit}). 
This corresponds to a separation of $\sim$ 8 pixels so is not an artefact caused by shifts in the 
echelle over periods of several hours. However, the signal-to-noise 
of the data is low, and follow-up observations are required to confirm this 
finding. We note that although the observations of Meyer \& Lauroesch (1999) may 
have also been expected to find such velocity substructure at their resolution of 
14 km\,s$^{-1}$, their observations were in the Na\,{\sc i} D lines, which 
in some cases do not show velocity components, even when the Ca\,{\sc ii} lines 
do (Ryans et al. 1996). 

The current observations could also be used to estimate the IVC Ca\,{\sc ii} column 
densities along the slit as explained in section \ref{lsana}. 
Figure \ref{elf_fits} displays the equivalent widths and Ca\,{\sc ii} column densities 
for a number of the slit positions for 
the second slit orientation. Our Ca\,{\sc ii} K equivalent widths 
range from 0.06$-$0.09 \AA. These are slightly lower than those measured by Lehner 
et al. (1999) towards other parts of the IVC with low-resolution data, 
which are in the range 0.05$-$0.20 \AA, with a median value of 0.10 \AA.  
Towards the nearby halo star HD 203664 (which has an IVC H\,{\sc i} column density 
of less than $\sim$ 10$^{18}$ cm$^{-2}$), Ryans et al. (1996) measured an IVC 
Ca\,{\sc ii} equivalent width of 0.06$\pm$0.01 \AA. 
The Ca\,{\sc ii}/H\,{\sc i} ratio for the current data towards the M\,15 IVC  
varies from 2.1$\times$10$^{-8}$ to 2.9$\times$10$^{-8}$. 
Unfortunately, poor quality of our measurements caused by low signal-to-noise, and 
uncertainties in determining the sky and continuum levels, results in our errors being 
larger than the differences between these values. 

\subsection{Wisconsin H$\alpha$ Mapper (WHAM) Results}

\subsubsection{The main IVC towards M\,15}

\begin{figure*}
\begin{center}
\epsffile{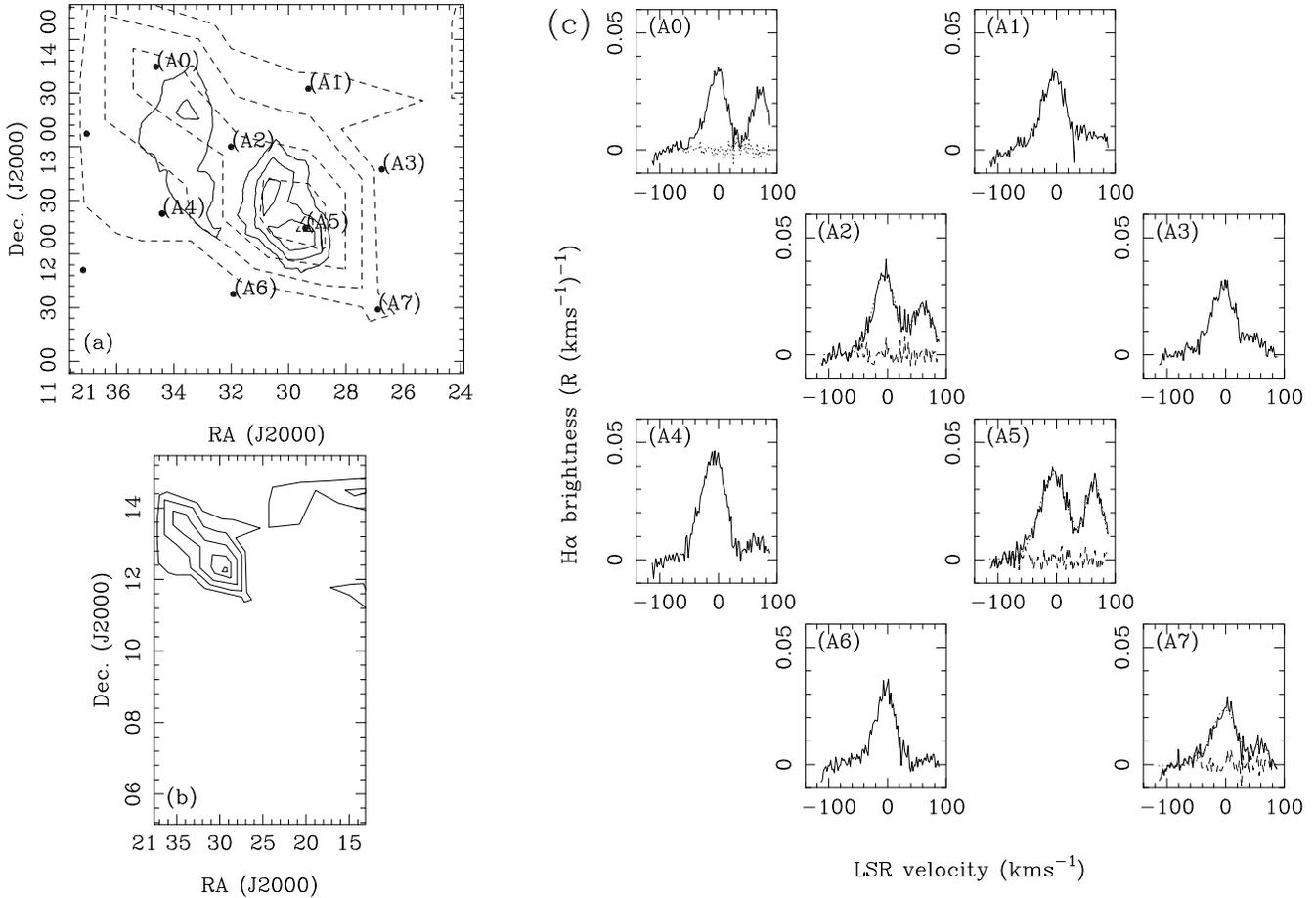}
\end{center}
\caption{(a) Lovell-telescope IV H\,{\sc i} column density map taken 
from Kennedy et al. (1998) in the vicinity of M\,15 with WHAM-observed positions 
marked (A0)$-$(A7). Contour levels (full line) are at N$_{HI}$=(1,2,3)$\times$10$^{19}$ cm$^{-2}$. 
The dashed lines show the H$\alpha$ contours from the WHAM H$\alpha$ observations, integrated from 
$\sim$ +50$-$90 km\,s$^{-1}$ with levels at 0.1$\times$(1,2,3,4,5) R. 
(b) H$\alpha$ brightness integrated from $\sim$ +50$-$90 km\,s$^{-1}$ for the whole field observed 
with the Lovell telescope in H\,{\sc i}. Contour levels are at 0.1$\times$(1,2,3,4,5) R.
(c) WHAM H$\alpha$ spectra at the positions (A0)$-$(A7). Velocities are in the LSR. Where 
H$\alpha$ is observed (positions A0,A2,A5,A7), a three-component Gaussian fit is indicated by a 
dotted line with the (data$-$model) residual shown by the dashed line.
}
\label{wham_m15_two}
\end{figure*}

Of the 184 spectra extracted from the WHAM survey, a clearly-defined 
H$\alpha$ component at intermediate velocities exceeding +50 km\,s$^{-1}$ is only obvious at 
three positions (A0, A2, A5) with a weak detection towards (A7). We note that because M\,15 itself 
is at an LSR velocity of $\sim -$100 km\,s$^{-1}$, it does not contaminate the observed 
spectrum. The spectra observed at these positions are shown in  
Fig. \ref{wham_m15_two}(c). Fig. \ref{wham_m15_two}(a) depicts the locations of 
these WHAM pointings relative to the Lovell telescope H\,{\sc i} surface density map,  
with integration of velocities from $\sim$ +50$-$90 km\,s$^{-1}$ in the WHAM H$\alpha$ 
data also being superimposed. Fig. \ref{wham_m15_two}(a) also shows the 
WHAM data integrated from $\sim$ +50$-$90 km\,s$^{-1}$, over the whole field 
mapped by the multibeam observations. The gas with brightness $\sim$ 0.1 R 
is quite extended about the main IVC condensation, with an additional weak signal 
detected towards the tentative H\,{\sc i} detection `E' (see Fig.  \ref{mbhicol}). 

Results of three-component Gaussian fitting at the four positions are shown in Table \ref{whamresults}, 
uncorrected for Galactic extinction. These fits take into account the instrumental profile of 
the instrument, which is comprised of a 12 km\,s$^{-1}$ Gaussian with low-level wings superimposed. 
Although almost equally-well fitted using just two components (for the LV and IV gas), we 
chose to fit using three components due to the slight asymmetry in the LV component, 
most clearly seen towards position (A7). Table \ref{whamresults} also shows the central 
velocity and velocity widths observed in H\,{\sc i} towards the component (A5), obtained by 
smoothing the Lovell telescope multibeam data to a resolution of 1 degree.

\begin{figure}
\begin{center}
\epsffile{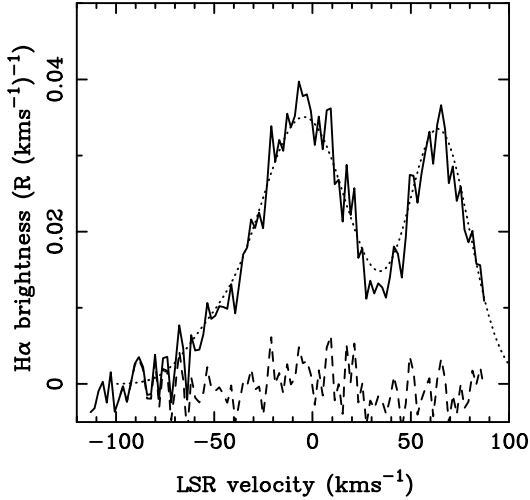}
\end{center}
\caption{Solid line: WHAM H$\alpha$ spectrum towards $l$=64.98$^{\circ}$, 
$b$=$-$27.16$^{\circ}$ (A5). The dotted line shows the results of a three-component Gaussian 
fit. The dashed line is a (data-three component Gaussian fit) residual. 
}
\label{wham_m15}
\end{figure}

The spectrum with the strongest IV H$\alpha$ emission (A5) is depicted in Fig. \ref{wham_m15}.  
This is the nearest grid position to the M\,15 IVC, whose centre lies approximately 
at $l$=65.01$^{\circ}$, $b=-$27.31$^{\circ}$. The peak IV H$\alpha$ 
brightness in this direction obtained using three-component Gaussian fitting is 0.035 Rayleighs 
(km s$^{-1}$)$^{-1}$ for the low velocity gas, and 0.033 Rayleighs (km s$^{-1}$)$^{-1}$ for the 
intermediate velocity gas, where 1 Rayleigh is 10$^{6}$/4$\pi$ photons cm$^{-2}$ s$^{-1}$ sr$^{-1}$. 
The integrated fluxes are 2.2$\pm$0.1 and 1.3$\pm$0.1 Rayleighs for the (total) LV and IV 
gas respectively. The Gaussian fit gives centroids of $-$53.9$\pm$4.6, $-$4.4$\pm$0.7 and 
+64.3$\pm$0.4 km\,s$^{-1}$, and FWHM velocity widths of 26.7$\pm$10.4, 47.8$\pm$2.6 and 
31.5$\pm$1.4 km\,s$^{-1}$, for the low and intermediate velocity gas respectively.  
The velocity centroid at this position (A5) agrees within the errors with the H\,{\sc i} 
data smoothed to the WHAM spatial resolution; this contrasts with the results 
of Tufte, Reynolds \& Haffner (1998) who tentatively found an offset in velocity between H$\alpha$ and 
H\,{\sc i} velocities of $\sim$ 10 km\,$^{-1}$ towards various HVC complexes. 
For the position (A5), the velocity width of the IV H$\alpha$ spectrum of $\sim$ 32 km\,s$^{-1}$ 
is some 10 km\,s$^{-1}$ greater than the H\,{\sc i} data at 1 degree resolution. For a 
gas at $\sim$ 10$^{4}$ K, some $\sim$ 22 km\,s$^{-1}$ of this is caused by thermal broadening, 
with the remaining width being due to non-thermal motions and beamsmearing of 
different components. The difference between the H\,{\sc i} and H$\alpha$ widths 
may imply that the two phases are not mixed. However, at least qualitatively, there is reasonable 
coincidence between the H$\alpha$ and H\,{\sc i} peaks (Fig. \ref{wham_m15_two}(a)), although 
the mapped area is small. 

The second positive detection (towards (A2) in Fig. \ref{wham_m15_two}), 
occurs in a region where the local H\,{\sc i} column density, at the Lovell telescope 
resolution of $\sim$ 12 arcmin, is lower than 1$\times$10$^{19}$ cm$^{-2}$. However, 
when smoothed to a WHAM resolution of 1 degree, the H\,{\sc i} column density 
at this point is $\sim$ 1.0$\times$10$^{19}$ cm$^{-2}$. 
Finally, towards (A7) there is a weak detection of $\sim$ 0.25 R at 
$\sim$ +59 km\,s$^{-1}$. At the WHAM resolution, the H\,{\sc i} column density 
at this point is $\sim$ 0.4$\times$10$^{19}$ cm$^{-2}$. We note that this position 
is close to the Arecibo-measured H\,{\sc i} position `AR' (Fig. \ref{mbhicol}), 
which has a similar velocity centroid of $v_{LSR}$=+61 km\,s$^{-1}$.  

\begin{table*}
\begin{center}
\small
\caption{Wisconsin H$\alpha$ Mapper results. The values for the velocities  
and fluxes are the results of three-component Gaussian fitting to the low and 
intermediate velocity gas. H$\alpha$ brightnesses have not been corrected 
for Galactic absorption, hence assuming that the IVC lies above the Galactic 
plane, the tabulated values are likely to be low by $\sim$ 25 per cent. 
Lovell telescope H\,{\sc i} multibeam results, smoothed to a WHAM-sized 1 degree 
beam, provide H\,{\sc i} velocity information towards position (A5). 
}
\label{whamresults}
\begin{tabular}{lrrrr}
Position on Fig. \ref{wham_m15_two}     
                           & (A7)                                          &
                             (A5)                                          &
                             (A2)                                          &
                             (A0)                                          \\
RA (J2000)                 & 21$^{h}$26$^{m}$53.3$^{s}$                    &                
                             21$^{h}$29$^{m}$24.9$^{s}$                    &
                             21$^{h}$32$^{m}$00.4$^{s}$                    &
                             21$^{h}$34$^{m}$37.0$^{s}$                    \\
Dec (J2000)                & +11$^{\circ}$28$^{\prime}$58$^{\prime\prime}$ &
                             +12$^{\circ}$14$^{\prime}$30$^{\prime\prime}$ & 
                             +13$^{\circ}$00$^{\prime}$02$^{\prime\prime}$ &
                             +13$^{\circ}$44$^{\prime}$43$^{\prime\prime}$ \\
$l$ (degrees)              &    63.88                                      &
                                64.98                                      &
                                66.09                                      &
                                67.19                                      \\
$b$ (degrees)              & $-$27.16                                      &
                             $-$27.16                                      &              
                             $-$27.16                                      &
                             $-$27.16                                      \\
H$\alpha$ LV $V_{c}$ (km\,s$^{-1}$)     & $-$32.0$\pm$16.6,  +2.7$\pm$2.6  &
                                          $-$53.9$\pm$4.6, $-$4.4$\pm$0.7  &
                                          $-$43.8$\pm$3.7, $-$5.3$\pm$0.6  &
                                          $-$32.4$\pm$9.5, $-$0.7$\pm$1.5  \\
H$\alpha$ LV FWHM (km\,s$^{-1}$)        &   44.6$\pm$22.4, 35.1$\pm$4.8    &
                                            26.7$\pm$10.4, 47.8$\pm$2.6    &
                                            18.8$\pm$9.4,  35.2$\pm$2.0    &
                                            30.3$\pm$14.1, 29.5$\pm$2.9    \\
H$\alpha$ LV Flux (R)                   &   0.27$\pm$0.18, 1.02$\pm$0.20   &
                                            0.14$\pm$0.06, 2.10$\pm$0.08   &
                                            0.11$\pm$0.04, 1.50$\pm$0.06   &
                                            0.20$\pm$0.13, 1.26$\pm$0.14   \\
H$\alpha$ LV Peak (mR (km\,s$^{-1}$)$^{-1}$)
                                        &4.8$\pm$1,     23$\pm$1 &
                                         4.2$\pm$1,     35$\pm$1 &
                                         4.7$\pm$01,    34$\pm$1 &
                                         5.2$\pm$1,     34$\pm$1 \\
H$\alpha$ IV $V_{c}$ (km\,s$^{-1}$) & +59.3$\pm$1.1  &
                                      +64.3$\pm$0.4  &
                                      +59.4$\pm$0.8  &
                                      +71.4$\pm$0.5  \\
H$\alpha$ IV FWHM (km\,s$^{-1}$)    & 18.8$\pm$3.4   &
                                      31.5$\pm$1.4   &
                                      38.1$\pm$2.4   &
                                      27.5$\pm$1.5   \\
H$\alpha$ IV Flux (R)               & 0.25$\pm$0.02  &
                                      1.30$\pm$0.04  &
                                      0.95$\pm$0.04  &
                                      0.91$\pm$0.03  \\
H$\alpha$ IV Peak (mR (km\,s$^{-1}$)$^{-1}$)
                                    &  11$\pm$1 &
                                       33$\pm$1 &
                                       20$\pm$1 &
                                       26$\pm$1 \\
H\,{\sc i} IV $V_{c}$ (km\,s$^{-1}$)    &  --             &
                                           65.7$\pm$0.6   &
                                           --             &
                                           --             \\
H\,{\sc i} IV FWHM (km\,s$^{-1}$)       &  --             &
                                           22.0$\pm$2.0   &
                                           --             &
                                           --             \\
\end{tabular}
\normalsize
\end{center}
\end{table*}

\subsubsection{WHAM pointings towards HD 203664 and position 'D'}

\begin{figure}
\begin{center}
\epsffile{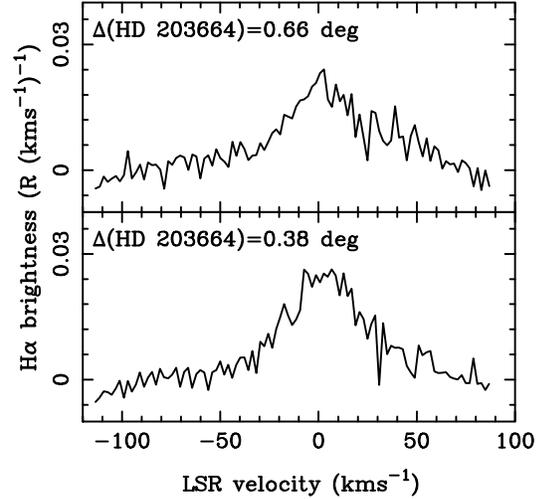}
\end{center}
\caption{Nearest two WHAM H$\alpha$ spectra towards HD 203664 with their angular 
distances from the star indicated. 
}
\label{hd203664_wham_m15}
\end{figure}

\begin{figure}
\begin{center}
\epsffile{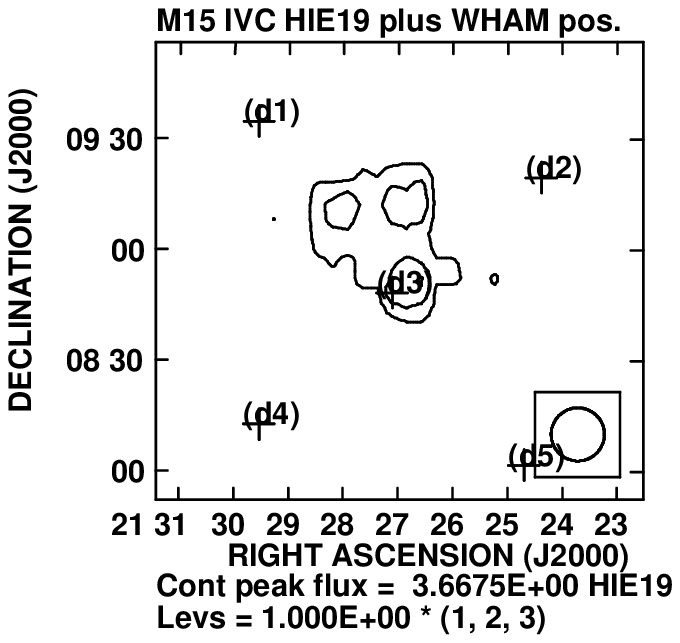}
\epsffile{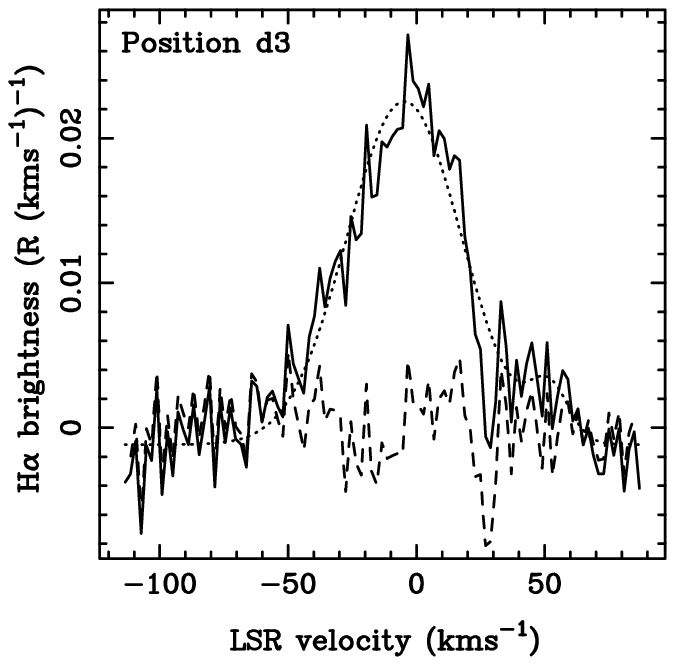}
\end{center}
\caption{Top panel: Lovell telescope multibeam IVC H\,{\sc i} surface 
density map of feature (D) (+50$< V_{LSR} <$+80 km\,s$^{-1}$), with contour levels at 
N$_{HI}$=(1,2,3)$\times$10$^{19}$ cm$^{-2}$ with WHAM pointings 
(d1)$-$(d5) marked. Bottom panel: WHAM H$\alpha$ spectrum towards (d3). The dotted 
line is a three-component Gaussian fit with the dashed line showing 
the data-fit residual spectrum. 
}
\label{posd_wham_m15}
\end{figure}

The nearest WHAM H$\alpha$ pointings in the vicinity of the halo star 
HD 203664 are some 0.38 and 0.66 degrees from the stellar position and are 
depicted in Fig. \ref{hd203664_wham_m15}. Although there is a slight excess of 
H$\alpha$ with velocities exceeding $\sim$ +50 km\,s$^{-1}$, this is very close 
to the noise and at lower velocities than the Ca\,{\sc ii} K absorption seen by Ryans et al. 
(1996) which had velocities of +75$-$80 km\,s$^{-1}$. 
Fig. \ref{posd_wham_m15} shows the WHAM pointings superimposed on the Lovell 
telescope multibeam H\,{\sc i} column density map for the IVC towards IVC position `D'  
(c.f. Fig. \ref{mbhicol}). Again, although there is some H$\alpha$ excess towards position (d3), 
this is very weak (brightness 0.07 R), and is also at a low-velocity of $\sim$ +50 km\,s$^{-1}$; 
this compares with the IVC H\,{\sc i} velocity of $\sim$ +68 km\,s$^{-1}$ at this point. 
Even though the WHAM pointing centre is just within the N$_{HI}$=1$\times$10$^{19}$ cm$^{-2}$ 
contour, the relatively small size of the `D' component results in the H\,{\sc i} column 
density at this point at 1 degree resolution being $\sim$ 0.5$\times$10$^{19}$ cm$^{-2}$. 

\subsection{IRAS ISSA results}

\begin{figure}
\begin{center}
\epsffile{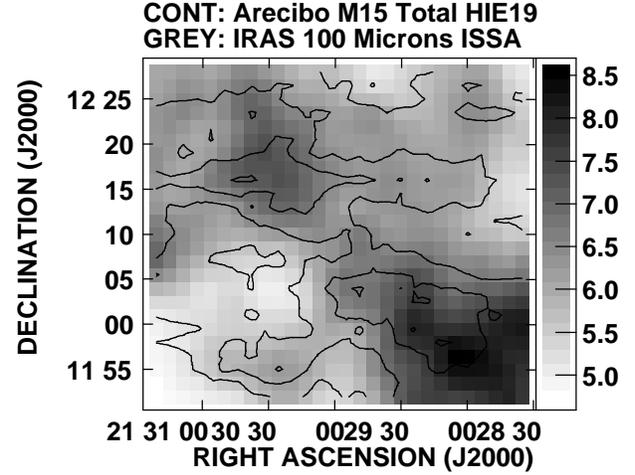}
\end{center}
\caption{Arecibo H\,{\sc i} and IRAS ISSA data. Greyscale: IRAS 100 micron flux density. 
Contour: Total (LV plus IV) H\,{\sc i} column density at 
N$_{HI}$=(56,60,64,68,72)$\times$10$^{19}$ cm$^{-2}$.
}
\label{hiiras100}
\end{figure}

\begin{figure}
\begin{center}
\epsffile{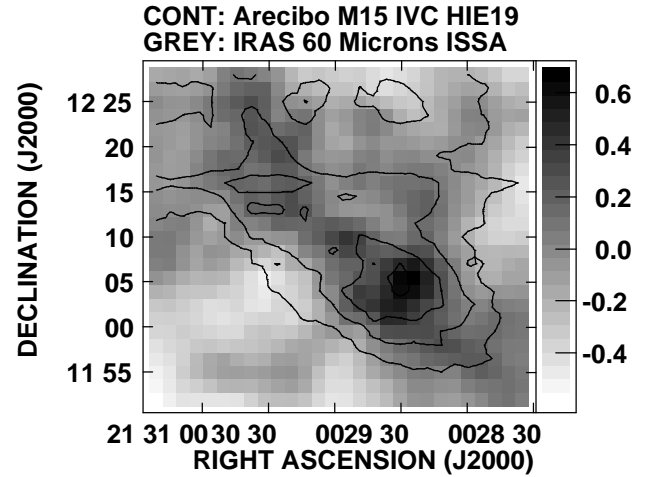}
\end{center}
\caption{Arecibo H\,{\sc i} and IRAS ISSA data. Greyscale: IRAS 60 micron flux density. 
Contour: IVC H\,{\sc i} column density at N$_{HI}$=(2,4,6,8)$\times$10$^{19}$ cm$^{-2}$.
Note how the peak H\,{\sc i} contour corresponds with the IRAS peak. 
}
\label{hiiras60}
\end{figure}

\begin{figure}
\begin{center}
\epsffile{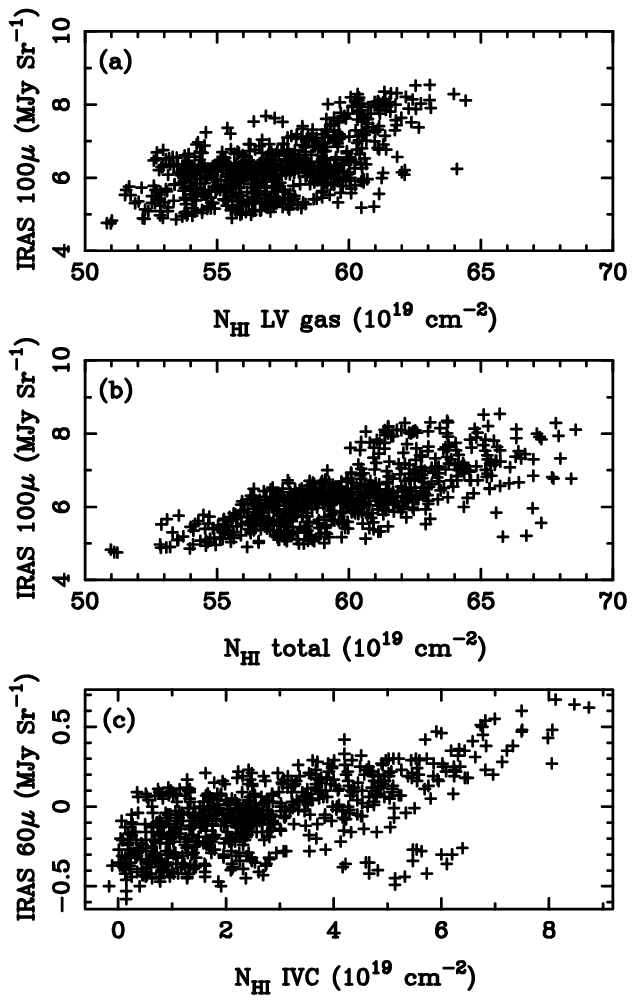}
\end{center}
\caption{Arecibo H\,{\sc i} and IRAS ISSA data. The IRAS fluxes have an 
arbitrary zero-point. (a) IRAS 100 micron flux density plotted against low velocity gas 
H\,{\sc i} column density. (b) IRAS 100 micron flux density plotted against total 
H\,{\sc i} column density. (c) IRAS 60 micron flux density plotted against IVC 
H\,{\sc i} column density alone.
}
\label{hiiras}
\end{figure}

In this section the IRAS 60 and 100 micron ISSA images are compared with previous 
Arecibo H\,{\sc i} observations of Smoker et al. (2001a). These latter data are of 
resolution 3 arcmin and contain both low and intermediate velocity gas. Fig. 
\ref{hiiras100} shows the 100 micron image overlaid on the total (LV plus IV) H\,{\sc i} 
column density, with Fig. \ref{hiiras60} displaying the corresponding 60 micron data and the 
IVC H\,{\sc i} column density alone. There appears to be a relatively good correlation between 
the IRAS and H\,{\sc i} results, both for the low and intermediate velocity gas; although with 
such a small field such agreement could have occurred by chance. The emission is not likely to 
be from the M\,15 globular cluster itself, as globular clusters are undetected at the 60 and 100 
micron IRAS sensitivity limits (Origlia, Ferraro \& Pecci 1996). Plots comparing the low, 
total and intermediate H\,{\sc i} column density with the IRAS flux density are displayed 
in Fig. \ref{hiiras}. Assuming that the observed correlations are in fact real, the 
current data indicate that this IVC contains dust, as has also been observed in other such 
objects (Malawi 1989; Wei$\beta$ et al. 1999). The fact that the IVC is tentatively detected 
at 60 microns indicates that the dust may be warm, perhaps heated by collisions. 
 
The (tentative) IRAS detection contrasts with the situation in HVCs, which at least at the 
IRAS sensitivity do not appear to emit at these wavelengths (Wakker \& Boulanger 1986). Such a 
difference indicates either differences in dust content, dust temperature or 
environment (such as distance from the heating field of the Galactic plane). 
H\,{\sc i} absorption line spectroscopy towards a number of HVCs by 
Akeson \& Blitz (1999) indicates that the fraction of cold gas is low in these 
objects. If so, then this would point to the Galactic-halo HVCs containing less dust 
than the IVCs and not differences in temperature.  

\section{Discussion}
\label{disc}

\subsection{The H$\alpha$ properties of the M\,15 IVC}

\subsubsection{Estimates of the emission measure and H\,{\sc ii} column density}

The detection of H$\alpha$ towards the main M\,15 IVC (position A5) 
with brightness of $\sim$ 1.3 R compares to lower values of between 0.06 to 0.20 R 
observed in a number of High Velocity Clouds (Tufte et al. 1998). 
Similarly, many of the HVCs observed by Weiner, Vogel \& Williams (1999) are very 
faint ($<$0.03 R), implying that, if they are photoionised by hard photons from the 
Galaxy, they are 20--60 kpc distant. The relatively high H$\alpha$ emission from the 
M\,15 IVC in turn argues for a closer distance, assuming that shock ionisation is not 
the dominant factor. This is a big assumption, as 
detections of H$\alpha$ in the Magellanic Stream (distance 55 kpc) of 
are 0.20$-$0.37 Rayleigh (Bland-Hawthorn \& Maloney 1999) are similar to the range 
of $\sim$ 0.1$-$0.5 Rayleigh detected towards the IVC Complex K 
(Haffner, Reynolds \& Tufte 2001a) which has a distance bracket of 0.3$-$7.7 kpc. 

Assuming that extinction is negligible and that the cloud is optically thin, 
the observed H$\alpha$ brightness of the IVC can be used to estimate both the column density 
of ionised hydrogen and the electron density. For an optically thin gas, the 
H$\alpha$ surface brightness in Rayleigh, $I_\alpha$, is given by;

\begin{equation}
I_{\alpha} = 0.36 \times {\rm EM} \times \sqrt{T} \times (1 - 0.37 \times {\rm ln}(T)),
\label{em1}
\end{equation}

\noindent
where $T$ is the gas temperature in units of 10$^{4}$ K and EM is the emission 
measure ($\int$n$_{e}^{2}dl$) in units of pc cm$^{-6}$ (Simonetti, Topasna \& 
Dennison 1996). Here, $n_{e}$ is the 
electron density and integration over $dl$ provides the thickness ($L$) of the ionised 
region. Assuming the temperature of the H\,{\sc ii} is 8,000 K, typical of the warm 
interstellar medium (Reynolds 1985), the resulting emission measure towards position 
(A5) is 3.7 pc cm$^{-6}$. 

Assuming a distance to the main IVC condensation of 1 kpc, the cloud size ($d$) at 
a column density limit of N$_{HI}=$10$^{19}$ cm$^{-2}$ is $\sim$ 0.6$^{\circ} 
\times$0.8$^{\circ}$, corresponding to linear dimensions of $\sim$ 10$\times$14 pc.
If the cloud is approximately spherically symmetric, this results in an estimated electron 
density (assuming a size of 12 pc) of $\sqrt{{\rm EM}/d}$ $\sim$ 0.6$f^{-1/2}$ cm$^{-3}$ 
and a column density of ionised hydrogen (=$L \times n_{e}$) of $\sim$ 2$\times$10$^{19} 
f^{1/2}$ cm$^{-2}$, where $f$ is the filling fraction of H\,{\sc ii} in the IVC. Smoothed to 
a resolution of 1 degree (corresponding to the size of the WHAM beam), the H\,{\sc i} column 
density at this point derived from the data of Kennedy et al. (1998) is 
$\sim$ 1.4$\times$10$^{19}$ cm$^{-2}$, hence the fractional ratio of 
H\,{\sc ii} to H\,{\sc i} is of order 1.4$\times f^{1/2}$, indicating that there is a 
substantial amount of ionised gas at intermediate velocities present in this sightline.

For positions (A2) and (A7), and using the same cloud size as above, results in 
estimated H\,{\sc ii} column densities of 1.5$\times$10$^{19}$ cm$^{-2}$ and 
0.8$\times$10$^{19}$ cm$^{-2}$, corresponding to fractional H\,{\sc ii} to H\,{\sc i} 
values of order 1.5$\times f^{1/2}$ and  2.0$\times f^{1/2}$ respectively. 
These fractional H\,{\sc ii} to H\,{\sc i} ratios towards the current IVC 
are somewhat higher than derived by Tufte et al. (1998) for a sample of HVCs 
which were calculated to be $<$ 0.06$\times f^{1/2}$. However, recent work by 
Bluhm et al. (2001), in sightlines towards the Large Magellanic Cloud, used the 
relative underabundance of neutral oxygen to infer an ionisation level in both an 
IVC and HVC of $\sim$ 90 per cent. It would be useful to observe this cloud 
using the same methods as described in the current paper and compare the 
results. 

Towards (d3) there is no detection in H$\alpha$ with velocities corresponding 
to the H\,{\sc i} value of $\sim$ +68 km\,s$^{-1}$. Assuming a cloud size of 8pc and
upper limit to the H$\alpha$ brightness of 0.2 R, gives an estimated upper limit 
to the IVC H\,{\sc ii} column density and fractional H\,{\sc ii} to H\,{\sc i} 
values of $\sim$ 4$\times$10$^{18}$ cm$^{-2}$ and 0.7$\times f^{1/2}$. Finally, 
we consider the halo star HD 203664, towards which the limiting column density of 
neutral hydrogen is $\sim$ 10$^{18}$ cm$^{-2}$. If we take the upper limit to the H$\alpha$ 
brightness of 0.2 R, and combine this with a cloud size $L$ (in pc, where the 
Arecibo beam is $\sim$ 1 pc FWHM at 1 kpc distance), a fractional 
H\,{\sc ii} to H\,{\sc i} ratio of $\sim$ 1.4$\times L \times f^{1/2}$ can be set 
by the current observations. 

\subsubsection{Estimates of the ionising radiation field and electron density}

Following Tufte et al. (1998), if photoionisation is the dominant cause of H$\alpha$ emission, 
then the incident Lyman continuum flux $F_{\rm LC}$ can be estimated thus, assuming case 
B recombination; 

\begin{equation}
F_{\rm LC} = 2.1 \times 10^{5} \frac{I_{\alpha}}{0.1 \rm{R}} \, {\rm photons \, cm^{-2} \, s^{-1}},
\label{halpha}
\end{equation}
-where $I_{\alpha}$ is the H$\alpha$ intensity in Rayleigh. For the M\,15 IVC, 
equation \ref{halpha} implies an incident flux ($F_{\rm LC}$) of 2.7$\times$10$^{6}$ 
photons cm$^{-2}$ s$^{-1}$. Hence if photoionisation is the main cause of H$\alpha$ production, 
the derived Lyman-alpha continuum flux towards the main M\,15 IVC condensation is a factor 
6$-$22 times higher than the implied incident flux towards the A, C and M HVCs observed 
by Tufte et al. (1998), and more than twice that observed towards the Complex K IVC 
by Haffner et al. (2001a). We recall that Complex A lies between 4$-$10 kpc, with 
Complex C being some 5$-$25 kpc distant and the M\,15 IVC being closer than 
3 kpc. In the future, comparison of derived Lyman-alpha continuum fluxes 
for a larger sample of IVC and HVC sightlines with known distances may provide information 
on the relative contributions of the Galactic and extragalactic ionising field. 

An alternative possibility is that the H$\alpha$ is produced by shock ionisation, 
caused by interaction of the IVC with LV gas. This is a real possibility given 
the orientation of the IVC and the fact that its $z$-distance of less than 
1 kpc puts it in the lower Galactic halo. 
Towards the nearby halo star HD 203664 in which IV absorption is seen, 
Sembach (1995) postulated that the dust grains in the IVC have been processed by such 
shocks which also currently produce the highly ionised species. For an ambient temperature 
of $<$3$\times$10$^{5}$ K, a cloud of velocity 50 km\,s$^{-1}$ will be supersonic and 
hence shocks may be formed given the right conditions. Using the models of Raymond (1979), 
which are applicable for shocks with speed 50$<$V$_{\rm s} <$140 km\,s$^{-1}$, the 
face-on H$\alpha$ surface brightness ($I_{\alpha perp}$) can be related to the number 
density of the pre-shocked gas ($n_{0}$), thus;

\begin{equation}
(I_{\alpha perp}) \sim 6.5 \times n_{0} (V_s/100)^{1.7} {\rm R}
\label{shockeq}
\end{equation}

Given that $I_{\alpha perp}$=1.3 R, and assuming a shock speed of 50 km\,s$^{-1}$, 
leads to an upper limit to $n_{0}$ of 0.6 cm$^{-3}$. The value is an upper limit (for this 
shock speed velocity) as a non-perpendicular sightline will increase the observed 
$I_{\alpha}$ (Tufte et al. 1998). Of course, given the fact that the transverse 
component of the velocity of the M\,15 IVC is unknown, this value is very uncertain. 
Discriminating between shock and photoionisation is difficult, although further 
progress may be possible via measurements of appropriate emission line ratios 
(Tufte et al. 1998). 

\subsection{Ca\,{\sc ii} number density towards the M\,15 IVC and HD 203664}

Before discussing the Ca\,{\sc ii} K results, we note that calcium is depleted onto dust and is 
not the dominant ionisation species, hence the absolute metallicity of the M\,15 IVC is uncertain 
and awaits high resolution UV observations. As emphasized by Ryans et al. (1997), differences 
in resolution between the optical and radio data, combined with the Ca\,{\sc ii} K 
results only placing limits on the ion abundances, makes it important not to over-interpret 
the observed N(Ca\,{\sc ii})/N(H\,{\sc i}) ratios. 

With that caveat, {\em and assuming that the current observations do not miss any 
narrow-velocity components}, the average IVC Ca\,{\sc ii} number density and 
ratio of the IVC compared to H\,{\sc i} towards the centre of M\,15 
are log$_{10}$(N(Ca\,{\sc ii} cm$^{-2}$)=12.0 and 
N(Ca\,{\sc ii})/N(H\,{\sc i}) = 2.5$\times$10$^{-8}$. The H\,{\sc i} 
column density towards the M\,15 centre is 4$\times$10$^{19}$ cm$^{-2}$ and 
was obtained using the combined WSRT plus Arecibo map of resolution 
111$^{\prime\prime} \times$56$^{\prime\prime}$. For the nearby sightline 
towards the halo star HD 203664, we use the results of Ryans et al. (1996) for the 
total IVC Ca\,{\sc ii} K column density of $\sim$1$\times$10$^{12}$ cm$^{-2}$, combined 
with the upper limit to the H\,{\sc i} column density at a resolution of 3 arcmin 
towards HD 203664 of $\sim$1$\times$10$^{18}$ cm$^{-2}$ (Smoker et al. 2001a), 
giving a much higher value of N(Ca\,{\sc ii})/N(H\,{\sc i}) $>$ 1.0$\times$10$^{-6}$. 

The current results compare with literature values of N(Ca\,{\sc ii})/N(H\,{\sc i}) $\sim$ 
2$\times$10$^{-9}$ (Hobbs 1974, 1976) for low velocity diffuse clouds and of 
N(Ca\,{\sc ii})/N(H\,{\sc i}) in the range 3--300$\times$10$^{-9}$ cm$^{-2}$ for 
the high latitude clouds studied by Albert et al. (1993). The N(Ca\,{\sc ii})/N(H\,{\sc i}) 
ratios in IVCs and HVCs are thought to be higher than in low velocity gas due to the former  
having less dust onto which calcium is depleted. Thus the observed N(Ca\,{\sc ii})/N(H\,{\sc i}) 
ratio of 2.5$\times$10$^{-8}$ (or $\sim$ 0.01 of the total solar calcium abundance) 
in the line-of-sight towards M\,15 is typical of other high latitude clouds and also of 
other HVCs and IVCs (e.g. Wakker et al. 1996a; Ryans et al. 1997).

The lower limit of N(Ca\,{\sc ii})/N(H\,{\sc i}) $>$ 10$^{-6}$ towards HD 
203664 is, however, on the high side for IVCs/HVCs, being $\sim$ 0.5 of 
the total solar calcium abundance. There are several possible reasons for this. 
Firstly, the (currently undetected) H\,{\sc i} towards HD 203664 could be 
in a clump of gas smaller than the Arecibo beamsize of 3 arcmin; if this were 
the case then the H\,{\sc i} column density limit used would be too low
and the derived N(Ca\,{\sc ii})/N(H\,{\sc i}) ratio too high. 
Additionally, HST UV observations towards HD 203664 indicate that the H\,{\sc i} 
towards this object is at least partially ionised (Sembach, private communication), 
either by shock ionisation or photoionisation. If photoionisation, aside from the normal 
ionising source being Galactic OB-type stars or the extragalactic UV field, HD 203664 
itself (spectral type B0.5) could be a possible ionising source. The  
fact that its LSR velocity is +110 km\,s$^{-1}$ (Little et al. 1994) compared with 
the IVC at +70 km\,s$^{-1}$ is inconclusive in determining the relative distance 
of the line of sight IVC towards HD 203664 with the star itself. Finally, there 
remains the possibility that the hydrogen is in molecular form. However, given 
the low H\,{\sc i} column density towards the HD 203664 sightline, this appears 
unlikely. 

Alternatively, it could be that the derived value of N(Ca\,{\sc ii})/N(H\,{\sc i}) 
$>$ 10$^{-6}$ towards HD 203664 is correct. This would tally with 
the IUE results of Sembach (1995), which found that the majority 
of the elements in the IV gas, when referenced to sulphur, were within a factor 
5 of their solar values and strongly point to a Galactic origin for this part of the
IVC. The fact that our derived value for N(Ca\,{\sc ii})/N(H\,{\sc i}) towards the M\,15 IVC 
of 0.01 solar is much lower than towards HD 203664 is likely to be caused by different 
ionisation fractions and dust contents, and/or differing formation mechanisms. 
Clearly the latter is speculative and requires follow-up high resolution UV observations 
towards the M\,15 IVC to determine the abundances of elements such as sulphur and zinc that 
are not depleted onto dust. 

\subsection{Velocity widths and temperatures towards the M\,15 IVC and HD 203664}

Towards the main M\,15 IVC condensation, (feature `A' on Fig. \ref{mbhicol}) 
values of the H\,{\sc i} FWHM velocity width at resolutions of $\sim$2$\times$1 arcmin 
range from 5$-$14 km\,s$^{-1}$, corresponding to maximum kinetic temperatures 
in the range $\sim$500$-$4000 K. Mid-way between the M\,15 IVC and HD 203664, 
the FWHM equals 12 km\,s$^{-1}$ at a resolution of 3 arcmin, which corresponds  
to T$_{k}$ $\sim$ 3000 K (feature `AR' on Fig. \ref{mbhicol}). The current 
observations have additionally observed feature `D' at a resolution 12 arcmin, 
with FWHM velocity width also of 12 km\,s$^{-1}$, indicating gas of similar temperature. 
We note that each of these 
temperatures will be upper limits due to beamsmearing and turbulent velocity 
components. Finally, towards HD 203664, Ryans et al. (1996) found cloudlets with 
FWHM of 2.8 and 3.2 km\,s$^{-1}$ in Ca\,{\sc ii} K, corresponding to upper limits for the 
kinetic temperatures of $\sim$ 8000$-$10,000 K. Towards the same star, 
Sembach (1995) used the relative abundances of low-ionisation species to derive a 
temperature for the HD 203664 IVC of 5300$-$6100 K. 

The higher IVC gas temperatures towards HD 203664 than towards the M\,15 IVC, 
feature `D' or the intermediate position `AR' could be interpreted as being caused 
by the former cloud being nearer to the ionising field of the Galactic plane than the 
other two IVCs (Lehner 2000). It seems more likely that the lower 
temperatures seen towards parts of `A' and `AR' are simply caused by shielding of 
parts of these gas clouds; shielding that is not possible towards the HD 203664 
IVC because of the lower gas density there. A two-phase core-envelope structure 
for halo HVCs {\em has} often been proposed within the Galactic corona for  
1$<z<$5 kpc (e.g. Wolfire et al. 1995), where the two components of 100 and 
10,000 K are entrained by pressure from the hot Galactic corona. We note that 
the high-end H\,{\sc i} temperatures observed towards the M\,15 IVC indeed occur towards its 
outer parts where the H\,{\sc i} column density is low and the FWHM velocity widths 
are uncertain.  

\subsection{Comparison of H\,{\sc i} properties with previous Galactic halo IVCs and HVCs}

In this section we compare the high-resolution H\,{\sc i} 
properties of the M\,15 IVC with other IVCs and HVCs known to lie within the Galactic 
halo and observed at comparable resolution. We note that it is likely
that there are many different types of HVC, with recent work, for example, indicating that a 
number of the compact HVCs lie at distances of several tens of 
kpc (Braun \& Burton 2000). A literature search found the following objects
with known distances and observed in H\,{\sc i} at high resolution:  
the M\,13 IVC (Shaw et al. 1996), the 4-Lac HVC 100--7+100 (Stoppelenburg 
et al. 1998), the Low-Latitude Intermediate Velocity Arch (Wakker et al. 1996b), IVC 135+54--45 
(Wei$\beta$ et al. 1999), HVC 132+23$-$211 (within Complex A; Schwarz et al. 1976) 
and the M\,92 HVC (within Complex C; Smoker et al. 2001b). 

Table \ref{comptab} compares the properties of these IVC and HVC H\,{\sc i} gas clouds known 
to exist in the Galactic halo. Inspecting the table, there are no clear differences 
in the H\,{\sc i} properties (column density, peak temperature) of the two types of objects, 
which show a large scatter both within IVCs and HVCs. Similarly, the velocity widths 
of all of the objects, barring the peculiar HVC100$-$7 and the Mirabel cloud, all show minimum 
values of FWHM $\sim$ 5 km\,s$^{-1}$ and indicating that gas of temperature less than 
$\sim$ 500 K is common in these objects. This is in contrast to the situation observed 
at lower resolution for some northern clouds, where IVCs tend to have smaller velocity 
widths than their HVC counterparts (Davies, Buhl \& Jafolla 1976). 

If some IVCs and HVCs are formed from infalling gas, sweeping up high-$z$ H\,{\sc i} as 
they fall towards the plane, or if they are objects within the plane, or if they are formed 
within a Galactic fountain, it seems plausible that they could be preferentially 
aligned with the Galactic plane. Of course, it must be taken into consideration that 
at high resolution, only small parts of cloud complexes are studied, and by chance, 
some of these components will be aligned with the plane. In any case, 
of the four objects in the sample with a well-defined axis and at high Galactic latitude, 
three have their major axis near-parallel with the plane. 
Although there exist a number of such objects observed at lower resolution with this 
orientation, to our knowledge no systematic survey has been performed determining 
the orientation parameters of HVCs. If performed, this could act as a further discriminator 
between HVCs known to exist in the Galactic halo, and the sample of HVCs postulated to 
lie at extragalactic distances. 

Summarising, at present there are too few high-resolution observations of Galactic 
halo IVCs and HVCs to determine differences in H\,{\sc i} properties and any 
relationship between the two types of object. However, as previously noted, 
the IRAS and H$\alpha$ properties {\em do} appear to differ, although the 
number of objects studied in all three wavebands remains small.  

\begin{table*}
\begin{center}
\small
\caption{Previous high-resolution H\,{\sc i} observations of IVCs and HVCs known to exist in the 
halo, compared with the current results. T$_{p}$ and N$_{HI}$ are the peak temperatures 
in K and peak H\,{\sc i} column densities in units of 10$^{19}$ cm$^{-2}$ respectively. 
$\Delta$V1 and $\Delta$V2 are the broad and narrow velocity widths in km\,s$^{-1}$. The 
field ``Parallel to the Gal. plane' only refers to the object observed at high-resolution and 
not to the whole complex in the case of the large HVCs.
References; RKSD97, Ryans et al. (1997); 
SSH76, Schwarz et al. (1976); 
SBK96, Shaw et al. (1996); SSW98, Stoppelenburg et al. (1998); 
SKD01, Smoker et al. (2001b);  WHS96, Wakker et al. (1996b); 
WPS99, van Woerden et al. (1999); 
WWHM99 Wei$\beta$ et al. (1999). GSR velocities are calculated 
via V$_{GSR}$=V$_{LSR}$ + 250$\times$sin($l$)$\times$cos($b$) km\,s$^{-1}$.
}
\label{comptab}
\begin{tabular}{lrrrrrrrrrl}
 Cloud            &$l^{\circ}$ &     D      &  V$_{LSR}$  &$\Delta$V1    &    T$_{p}$  & IRAS? & Ca\,{\sc ii}/H\,{\sc i} & Parallel to & Res.               \\ 
 Ref.             &$b^{\circ}$ & (kpc)      &  V$_{GSR}$  &$\Delta$V2    &  N$_{HI}$   &       &                       & Gal. plane?   & $^{\prime}$        \\
                  &            &            &                 &          &             &       &                       &               &                    \\
 HVC 132+23$-$211 &    123     &  4$-$10    &     $-$211      & 5$-$15   &      25     &  N    & --                    &      Y        &  $\sim$ 2          \\
 (SSH76,WPS99)    &    +23     &            &     $-$329      &          &      26     &       &                       &               &                    \\
 M13SE IVC        &     59     &   $<$7     &       --73      & 27--35   &     0.7     &  N    & $>$1$\times$10$^{-7}$ &      N        &  $\sim$ 3$\times$2 \\
 (SBK96)          &    +41     &            &        +89      &      5   &     4.7     &       &                       &               &                    \\
 M\,15 IVC        &     65     &   $<$3     &        +70      & 5$-$15   &       8     &  Y?   & 2.5$\times$10$^{-8}$  &      Y        &  $\sim$ 2$\times$1 \\
 (This paper)     &  $-$27     &            &       +272      &     --   &      15     &       &                       &               &                    \\
 M92N HVC         &     68     &  5$-$25    &      --101      &  4$-$7   &     3.4     &  N    & --                    &      Y        &  $\sim$ 6$\times$6 \\
 (SKD01)          &    +34     &            &        +91      &     --   &       6     &       &                       &               &                    \\
 HVC100$-$7+100   &    100     & $<$1.2     &       +106      &     13   &     0.5     &  N    & --                    &      N        &  $\sim$ 2$\times$2 \\
 (SSW98)          &   $-$7     &            &       +350      &     --   &    0.16     &       &                       &               &                    \\
 IVC 135+54$-$45  &    135     &0.29$-$0.39 &       --45      &  4$-$5   &    $>$7     &  Y    & --                    &     --        &  $\sim$ 1$\times$1 \\
 (WHHM99)         &    +54     &            &        +59      &     --   &      30?    &       &                       &               &                    \\
 PG0859+593 LLIVC & +156.9     & 1.7$-$4.0  &       --51      &    +19   &     2.0     &  ?    & --                    &     --        &  $\sim$ 2$\times$2 \\   
 (RKSD97, WHS96b) &  +39.7     &            &        +24      &          &     5.3     &       &                       &               &                    \\
 PG0906+597 LLIVC & +156.2     & 1.7$-$4.0  & --48, --53      &  28, 6   & 1.6, 3.9    &  ?    & --                    &     --        &  $\sim$ 2$\times$2 \\         
 (RKSD97, WHS96b) &  +40.6     &            &  +29,  +24      &          & 11.0, --    &       &                       &               &                    \\

\end{tabular}
\normalsize
\end{center}
\end{table*}

\section{Summary and Conclusions}
\label{concl}

The current H\,{\sc i} WSRT synthesis observations have shown that on scales down 
to $\sim$ 1 arcmin, the M\,15 IVC shows substructure, with variations in the column 
density of a factor of $\sim$ 4 on scales of $\sim$ 5 arcmin being observed, 
corresponding to scales of $\sim$ 1.5 $D$ pc, where $D$ is the the IVC distance in 
kpc. Of course, this is not an unexpected finding, but once again demonstrates that 
great care must be taken in interpreting quantities such as cloud metallicities which 
are derived from a combination of low-resolution radio plus optical data. The 
Lovell telescope H\,{\sc i} observations towards this cloud demonstrated how 
relatively large areas of sky can be mapped with the multibeam system 
in a short period of time in the search for IVCs and HVCs. These data showed 
that the M15 IVC has components spread out over several square degrees, with component 
`D' being mapped for the first time at medium resolution (12 arcmin) and having a similar 
column density to the IV gas centred upon M15 itself. Both the H\,{\sc i} emission-line 
and Ca\,{\sc ii} absorption-line data showed {\em tentative} evidence for velocity substructure, 
perhaps indicative of cloudlets. The Ca\,{\sc ii}/H\,{\sc i} value of $\sim$ 2.5$\times$10$^{-8}$ 
towards the main M\,15 condensation is similar to that previously observed in other 
IVCs and HVCs. Towards HD 203664, the {\em observed} lower limit of 10$^{-6}$ is somewhat 
higher, although this may be caused by factors such as the H\,{\sc i} beam being 
unfilled or partial ionisation of the gas on this sightline. The H\,{\sc i} properties 
of the M\,15 IVC are indistinguishable from HVCs, although with the lack of distance information 
towards most HVCs, comparisons are difficult. 

The tentative detection of infrared emission from the M15 IVC, as in other IVCs, 
does distinguish it from HVCs, and either points to the M15 IVC containing more dust, and/or being 
closer to the heating field of the Galactic plane than HVCs, which as a class of 
objects are not detected in the IRAS wavebands. Similarly, the relatively strong 
H$\alpha$ emission (exceeding 1 Rayleigh) towards parts of the M15 IVC, {\em if} caused 
by photoionization, may place it closer to the Galaxy than HVCs. Again, however, this 
finding is uncertain due to the problem in distinguishing photoionisation from shock 
ionisation, uncertainties in dust content, and differences in H\,{\sc i} volume densities 
in different objects studied thus far. 

Future work towards this cloud should include higher-signal-to-noise 
observations in the Ca\,{\sc ii} line in order to determine if the cloud velocity 
substructure tentatively found in the current observations is in fact real, and 
whether the Ca\,{\sc ii}/H\,{\sc i} ratio determined by the current observations 
is lower than towards the HD 203664 sightline. This 
should be combined with $^{12}$CO(1--0) sub-mm observations in order to determine if 
molecular material exists towards the peaks in H\,{\sc i} column density and out 
of which stars may form. The determination of the falloff in H\,{\sc i} column density, 
of the cloud to low column density limits would also indicate the ionisation properties of the 
object and whether or not there is any interaction between the M\,15 IVC gas and low velocity 
material. Finally, UV observations towards M\,15 globular cluster stars, although difficult, would 
provide important information on the absolute metallicity of the gas towards this object 
for comparison with the HD 203664 sightline. 

\section*{acknowledgements}

We would like to thank the Netherlands Foundation for Research in Astronomy (NFRA) 
which is a national facility supported by the Netherlands Organization for Scientific 
Research. In particular, we are very grateful for the help that Robert Braun provided 
with the reduction of the WSRT data. 
JVS acknowledges NFRA for use of data reduction facilities and hospitality 
and to the staff of Jodrell Bank observatory for help with the Lovell telescope multibeam 
observations. We would particularly like to thank Chris Jordan, Rob Lang, 
Peter Boyce and Robert Minchin for help with the LT observations. JVS would also 
like to thank the staff of the William Herschel Telescope, 
which is part of the Isaac Newton Group of telescopes, La Palma. The Digitized Sky Surveys were 
produced at the Space Telescope Science Institute under U.S. Government grant NAG W-2166. 
The National Geographic Society - Palomar Observatory Sky Atlas (POSS-I) was made by the 
California Institute of Technology with grants from the National Geographic Society. 
IPAC is operated by the Jet Propulsion Laboratory (JPL) and California Institute of Technology 
(Caltech) for NASA. IPAC is funded by NASA as part of the IRAS extended mission program under 
contract to JPL/Caltech. The WHAM project is funded primarily through grants from 
the National Science Foundation with additional support provided by the University of 
Wisconsin. JVS also thanks Andrew George and PPARC for financial support and to 
the referee, Dr V. Kilborn for useful comments.


{}

\end{document}